\documentclass[prd,amssymb,twocolumn,nofootinbib,10pt]{revtex4-2}

\pdfoutput=1
\usepackage{amsmath,amsfonts,amssymb}
\usepackage{color,verbatim,graphicx,float}
\usepackage{bbold}
\usepackage[dvipsnames]{xcolor}

\usepackage{latexsym}
\usepackage{rotating}
\usepackage{dsfont}

\usepackage{color}
\usepackage[active]{srcltx}
\usepackage{amsmath,amsfonts,amssymb,amsthm,amstext,amscd,eucal,srcltx}
\usepackage{epsfig,graphicx,bm}
\usepackage{epstopdf,epsf}
\usepackage{dcolumn}
\usepackage{hyperref}

\begin{document}

\small{\textsc{Version accepted for publication in Phys. Rev. D.}}\normalsize

\vspace*{3mm}
\title{General-relativistic spin system}

\author{${}^{\hslash, c}$Danilo Artigas}
\author{${}^{G}$Jakub Bilski}
\author{${}^{\hslash}$Sean Crowe}
\author{${}^{\hslash}$Jakub Mielczarek}
\author{${}^{\hslash}$Tomasz Trze\'sniewski}

\affiliation{${}^\hslash$ Institute of Theoretical Physics, Jagiellonian University, \L ojasiewicza 11, 
30-348 Cracow, Poland, EU \\
${}^c$ Universit\'e Paris-Saclay, CNRS,  Institut d'astrophysique spatiale, 91405, Orsay, France, EU \\
${}^G$Institute for Theoretical Physics and Cosmology,
Zhejiang University of Technology, \\ 310023 Hangzhou, China}

\begin{abstract}
\noindent
The models of spin systems defined on the Euclidean space provide powerful machinery for studying a broad 
range of condensed matter phenomena. While the nonrelativistic effective description is sufficient for most 
of the applications, it is interesting to consider special and general relativistic extensions of such models. 
Here, we introduce a framework that allows us to construct theories of continuous spin variables on a curved 
spacetime. Our approach takes advantage of the results of the nonlinear field space theory, which shows 
how to construct compact phase space models, in particular for the spherical phase space 
of spin. Following the methodology corresponding to a bosonization of spin systems into the spin wave 
representations, we postulate a representation having the form of the Klein-Gordon field. This representation 
is equivalent to the semiclassical version of the well-known Holstein-Primakoff transformation. The general-relativistic 
extension of the spin wave representation is then performed, leading to the general-relativistically motivated 
modifications of the Ising model coupled to a transversal magnetic field. The advantage of our approach is its 
off-shell construction, while the popular methods of coupling fermions to general relativity usually depend on the 
form of Einstein field equations with matter. Furthermore, we show equivalence between the considered spin system and the 
Dirac-Born-Infeld type scalar field theory with a specific potential, which is also an example of $k$-essence theory. 
Based on this, the cosmological consequences of the introduced spin field matter content are preliminarily investigated.   
\end{abstract}

\maketitle

\section{Introduction} \label{sec:I}
\noindent
The models of spin systems, for instance the Ising model, Heisenberg model, XY model \cite{Lieb}, or Hubbard model \cite{Hubbard},
provide a theoretical description of such phenomena as magnetism (ferromagnetism and antiferromagnetism), superconductivity,
and topological phase transitions \cite{Kosterlitz:1973xp}. While these models are adequate in the context of the ground-based tabletop
experiments, where local Euclidean geometry is the precise approximation, their extensions to a general curved spacetime are
almost unknown. Due to the weakness of gravity in the short-distance interactions, it is mostly irrelevant in condensed matter
physics\footnote{There are, however, some exceptions -- such as gravitational effects on Bose-Einstein condensate, employed
in ultraprecise gravimetry \cite{Muntinga:2013pta}. In this case, the phase of the wave function of a coherent many-body state
is affected by the gravitational potential of an external mass, such as Earth.}. In consequence, surprisingly little is known about the
condensed matter phenomena in curved spacetime. Although rather no one would ask the question about what would happen with a
ferromagnet during cosmic inflation, the lack of the definite answer reflects theoretical deficiencies in the lattice spin models. Analysis
of their relativistic extensions may not only bring us to the better fundamental understanding of the interaction between gravity and
spins but also provide theoretical foundations for possible future condensed matter experiments on Earth's orbit\footnote{First 
space experiments of this kind have already been performed \cite{Becker,Aveline:2020ob}.}. More abstract thought experiments, 
like the analysis of a superfluid state in the vicinity of a black hole horizon or the measurement of gravitational waves near a merger 
of black holes propagating through a spin glass, might be pondered in light of our results as well.

In the case of spin models describing the ground-based laboratory experiments, spins are considered to
be attached to given space points, i.e. forming a fixed lattice. Consequently, such models explicitly break
general covariance by distinguishing a certain reference frame. A useful step towards a frame-independent formulation
is the continuous limit of a spin system, in which a field-theoretic description of spins is obtained. The analogous continuous
spin system approximations play a powerful role in theoretical condensed matter physics, e.g. in the theory
of topological phase transitions \cite{Kosterlitz:1973xp}.

The system in such a case becomes a spin field $\vec{S}({\bf x})$ defined on some spatial hypersurface $\Sigma$.
In the standard condensed matter considerations, the spatial manifold is chosen as $\Sigma = \mathbb{R}^d$, 
where its dimension $d$ is 1, 2, or 3. From this perspective, one could naively presume that the desired 
generalization of a spin system is provided by the relativistic field theory of the vector field $\vec{S}({\bf x})$, with 
the constraint on its norm (the spin magnitude) 
$\forall{\bf x}: ||\vec{S}({\bf x})|| = \sqrt{\vec{S}({\bf x})\!\cdot\!\vec{S}({\bf x})} =: S$. 
Such a framework, an example of which is the famous nonlinear $\sigma-$model \cite{GellMann:1960np,Witten:1983ar}, 
however, does not lead to the correct theory of a spin field.

In our analysis, we identify the notion of a spin at each point of space with its semiclassical description by a vector 
$\vec{S}$ and initially restrict to the spatial manifold $\Sigma$ given by the three-dimensional Euclidean space. We are 
going to construct the relativistic extension of such a continuous distribution of spins. A natural, but naive, proposal for 
a generalization of a spin field on Euclidean space to either the spatial sector of spacetime or the Cauchy hypersurface 
constructed through the ADM decomposition does not lead to a well-working model\footnote{Actually, this proposal led
to the already mentioned nonlinear $\sigma-$model \cite{GellMann:1960np,Witten:1983ar}.}. The reason is that the 
spin field is not a standard classical tensor field.

In quantum physics, the model of spin is a finite-dimensional Hilbert space representation of the ${\rm SU}(2)$ group
of symmetries, i.e., rotations, and the $\mathfrak{su}(2)$ algebra of observables (whose generators are spin operators 
$\hat S^x$, $\hat S^y$, and $\hat S^z$). The $\mathfrak{su}(2)$ algebra is noncommutative, which leads to the uncertainty 
relations between the three spin operators and does not allow one to measure them simultaneously. One often considers the 
auxiliary object that is a vector $\vec{S}$ (called the spin vector), whose components are the expectation values of the 
$\hat S^x$, $\hat S^y$, and $\hat S^z$ operators in a given state. At the (semi)classical level, this vector spans 
the phase space ${\cal P}_S$, which is the two-sphere $\mathbb{S}^2$ of radius $S$, equipped with a natural symplectic 
form (cf. \cite{Nielsen:1988an,Alekseev:1988ql}). $\mathbb{S}^2$ does not have the structure of a cotangent bundle and, 
consequently, the decomposition of such a phase space into the product of configurational and momentum subspaces is 
possible only locally. The distinction between a configuration and a momentum of spin becomes clear in the context 
of the spin precession in a constant magnetic field, whose Hamiltonian we consider in this paper, since one angle on 
${\cal P}_S$ can be used to parametrize the circle circumscribed by the precessing spin, and the other will be a (constant) 
angle between the spin vector and the magnetic field \cite{Nielsen:1988an}. 
The interpretation of ${\cal P}_S$ as the phase space of spin can also be compared with models of a classical 
relativistic particle, where, depending on an approach, the phase space of spin in the rest frame is indeed 
${\cal P}_S$ \cite{Wiegmann:1989ms} or is obtained by constraining the product of momentum space (which is given by 
${\cal P}_S$) and configurational space (which is another $\mathbb{S}^2$) \cite{Rempel:2016is,Rempel:2017be} or by 
constraining the product of two unit $\mathbb{S}^2$ (i.e., configurations and momenta are described in the same way), 
with $S$ as a coupling constant \cite{Balachandran:1980sy}. All models that we mention here have their origin in 
applying the coadjoint orbit method \cite{Kirillov} to Poincar\'{e} group, and they differ only by a choice of phase 
space coordinates.

Nevertheless, in this paper, we do not investigate spinning particles but a continuous distribution of spin 
vectors $\vec{S} = \vec{S}({\bf x})$ (a normalized vector field), which is treated as a system consisting of spherical 
symplectic manifolds attached at each point of space\footnote{This might be considered as the continuous approximation 
of a spin lattice studied in condensed matter physics.}. Through a procedure of the boson mapping of operator 
algebras (bosonization)\footnote{The most popular bosonization methods are: 
the Holstein-Primakoff transformation \cite{Holstein:1940zp}, the Dyson-Maleev technique \cite{Dyson:1956zza,Dyson:1956zz,Maleev} 
and the Jordan-Schwinger map \cite{Jordan,Schwinger}.}, the degrees of freedom associated to a spin distribution 
can be represented by spin waves (a comprehensive description of this issue is given in \cite{Stancil}). Our particular choice 
is the classical analog of a spin wave reinterpreted as a bosonic field paired with its conjugate momentum. In this case, 
we will construct a scalar field representation, with the property that values of the field and its momentum are compactified 
to the sphere. Such a bosonic representation will allow us to construct the natural extension to a general-relativistic model. 
Translating it back to the spin vector's distribution is the main goal of our analysis, defining a method never (up to our knowledge) 
studied before. Our approach belongs to the recently introduced framework of nonlinear field space theories (NFSTs) 
\cite{Mielczarek:2016rax}, which extends the standard field theories to the case where the phase space of a field has the 
nontrivial topology. This generalization is related to numerous ideas and theories, including \emph{principle of finiteness} 
\cite{BornInfeld}, Born reciprocity \cite{Born:1938ay}, Relative Locality \cite{AmelinoCamelia:2011bm}, Metastring Theory 
\cite{Freidel:2015me}, and polymer quantization in loop quantum gravity \cite{Ashtekar_2003}. Some of the relations have 
been already discussed in detail in Refs. \cite{Mielczarek:2016rax,Mielczarek:2016xql,Bilski:2017gic,Trzesniewski:2017lpb}. 
It is worth noting that, as it has been suggested in the above mentioned literature, the compact phase space versions of 
NFSTs naturally implement the principle of finiteness of physical quantities, which underlies the Born-Infeld theory. However, 
the principle in the two theories is implemented in different manners, and a direct relation between the actions of Born-Infeld 
and NFST has not been found so far. Here, we show that the similarity between these theories is not only conceptual, and a 
scalar field with the specific relativistic spherical phase space reduces to Dirac-Born-Infeld (DBI)-type scalar field theory 
\cite{Leigh:1989jq,Silverstein:2003hf,Alishahiha:2004eh}.

The idea to relate spin systems and scalar fields is enabled by the equal dimensions of the local phase spaces of a 
spin system ($\mathbb{S}^2$) and a scalar field ($\mathbb{R}^2$)\footnote{The fact that the spherical $\mathbb{S}^2$
phase space has local $\mathbb{R}^2$ approximation has also been applied in the context of the gravitational phase 
space of the FLRW cosmological model \cite{Guimarey:2019lmn,Artigas:2020hfj}.}. Based on this observation, a 
new possibility of linking spin systems with field theories has been proposed in Ref.~\cite{Mielczarek:2016xql}. 
The exact matching of phase spaces (of a standard scalar field theory and a spin field model) is obtained by taking 
the large spin limit ($S \rightarrow \infty$). Following this reasoning, it has been shown that in the large spin limit 
the continuous Heisenberg XXX model is dual to the nonrelativistic scalar field theory with the quadratic dispersion 
relation \cite{Mielczarek:2016xql}. The result has been thereafter generalized to the Heisenberg XXZ model with 
a dimensionless anisotropy parameter $\Delta$. In this case, it has been demonstrated that taking both the large 
spin limit ($S \rightarrow \infty$) and the isotropic limit $\Delta \rightarrow 0$, we reduce the XXZ model to the
relativistic Klein-Gordon field \cite{Bilski:2017gic}. However, if the spin limit is not taken exactly, the next to the 
leading order terms violate relativistic symmetries \cite{Bilski:2017gic}. Therefore, the question is whether the 
construction can be improved to preserve the special relativistic and, furthermore, general relativistic symmetries, and
also for the spherical phase space field theories with an arbitrary value of the spin vector norm $S$. The purpose 
of this paper is to address this issue, directly constructing and analyzing the spin field theory that obeys general 
relativistic symmetries. What we do differently in the current paper is the relation between a spin field and a scalar 
field, which is now imposed in way equivalent -- at the level of excitations -- to the bosonization into spin waves.

The construction of the field formulation is based on the bosonization procedure of the fermionic interactions in a 
solid spin system. This procedure, first introduced in \cite{Tomonaga:1950zz}, aims to effectively describe the 
particle-hole-like excitations in the low-energetic regime \cite{Mattis:1964wp}. The known phenomenological realization 
of this method is the Tomonaga-Luttinger liquid model \cite{Tomonaga:1950zz,Luttinger:1963zz}, in which, under particular 
constraints, second-order interactions between electrons are represented by bosonic interactions. The model allows one to 
derive the exact spectrum of the Hamiltonian operator, free energy of noninteracting fermions, and dielectric constant 
\cite{Mattis:1964wp}\footnote{It is worth noting that Mattis and Lieb, who first presented these solutions, predicted vast 
applicability of the bosonization procedure, commenting on the Tomonaga-Luttinger model, ``We believe it has applications 
to the theory of fields which go beyond the study of the many-electron problem'' \cite{Mattis:1964wp}.}.

The basic excitations of coupled spin systems with fixed, homogeneous distribution of spin vectors (but not their 
orientations) are called spin waves (see \cite{Stancil} for a detailed introduction). One can distinguish two different 
kinds of such excitations. The first is described by the wave of deflected dipolar magnetic moments 
produced by elementary spins shifted from their equilibrium positions that propagates through the solid system. This 
type of excitations is significant at very long wavelengths, comparable to the spacing between individuals, and forms a 
macroscopic characteristic of the system. The second type of excitations relates to microscopic (quantum) properties 
of the spin lattice and is relevant for very short wavelengths, comparable with the lattice spacing.  
In both cases,  in the simplest realization, only the nearest-neighbor interactions are considered, forming a net 
of oscillators. 

Since we are interested in the continuous limit, in which the lattice spacing tends to zero, the first type of excitations
is relevant in our context. Furthermore, we are interested in the semiclassical description of the spin waves, understood 
as the linearly propagating vector's precession phase. When one considers closely spaced frequency components, they can 
be viewed as a wave packet that moves like a particle. This quasiparticle is called a magnon, and it does not obey the 
Pauli exclusion principle. We are going to describe properties of this bosonic field in the semiclassical picture, in which 
the spin variables are represented by the field variables corresponding to the matrix elements of the bosonic operators. 
The latter ones are constructed in the (semi)classical equivalent of the Holstein-Primakoff transformation \cite{Holstein:1940zp} 
(see Appendix~\ref{A} for details), which is one of the most popular maps selected for the bosonization procedure.

The paper is organized as follows. In Sec.~\ref{sec:II}, a spin-related spherical phase space and its scalar field-type 
parametrization are introduced. Then, in Sec.~\ref{sec:III}, a general strategy behind the defining of the 
Hamiltonian of the spin system is discussed, based on which, in Sec.~\ref{sec:IV}, the special relativistic spin 
system is introduced. The results are generalised to the curved spacetime case in Sec.~\ref{sec:V}. In Sec.~\ref{sec:VI}, 
the obtained model is shown to be be an example of the DBI-type $k$-essence model. Due to the cosmological relevance of 
the $k$-essence models, consequences of the considered field theory in the dynamics of Universe are preliminarily 
investigated in Sec.~\ref{sec:VII}. The results are summarized and additional discussion is given in Sec.~\ref{sec:VIII}.

\section{Phase space of a spin} \label{sec:II}
\noindent
As we mentioned in the Introduction, we will mostly restrict to the semiclassical description of spin by the vector 
$\vec{S} = (S^x,S^y,S^z)$, whose components are the generators of proper rotations in the three-dimensional Euclidean 
space (in other words, it is an element of the Lie algebra of rotations). Although the corresponding classical symmetry 
group is ${\rm SO}(3)$, we are more interested in another group, ${\rm SU}(2)$. The latter one is a double cover of the 
former; two-spheres are orbits of both of them, and their Lie algebras are isomorphic. The double cover property, which 
leads to the appearance of half-integer spins in the quantum theory, is the reason for selecting it to construct models 
of spin, as we do in this paper.

Another interesting property of spin is its dimension, $[M L^2 T^{-1}]$. It is the dimension of the classical angular 
momentum or the Planck constant, $\hbar$. It is worth noting that this is also the dimension of the physical action. 
The latter property already suggests that the geometrical interpretation of spin is linked with the phase space. 
In general, states of a classical spin (i.e., the intrinsic angular momentum assumed to be the classical 
counterpart of spin) are different directions in $\mathbb{R}^3$, equivalent to points on $\mathbb{S}^2$, and therefore, 
elements of such a space of states are naturally represented by two angles, $(\phi,\theta)$. As long as we do 
not define the associated conjugate momenta as belonging to some extra space (as in certain models of spinning 
particles mentioned in the Introduction), $\mathbb{S}^2$ spanned by $\vec{S}$ should actually be interpreted as the 
phase space of spin. The latter claim can be justified by applying the Kirillov orbit method \cite{Kirillov} to the 
group ${\rm SO}(3)$ (see \cite{Alekseev:1988ql}) or ${\rm SU}(2)$. The method allows one to construct a given phase space as a 
coadjoint orbit of a symmetry group $G$ (i.e. an orbit in $\mathfrak{g}^*$, e.g. $\mathfrak{su}(2)^* \cong \mathbb{R}^3$) 
for which the considered mechanical system remains invariant, while the corresponding quantum system should be described 
by an irreducible unitary representation of $G$. It is well known that two-spheres are coadjoint orbits of 
$G = {\rm SU}(2)$, as evidenced by the fact that a coset ${\rm SU}(2)/{\rm U}(1) \cong \mathbb{S}^2$ is the unit 
two-sphere. In consequence, the quantum models of spin are given by irreducible unitary representations of the 
${\rm SU}(2)$ group, labeled by half-integers $s = \frac{n}{2}$, where $n \in \mathbb{N} \cup \{0\}$, as expected.

In order to treat $\mathbb{S}^2$ as the phase space of the spin, one has to equip it with a symplectic form (a closed two-form) 
so that it becomes a symplectic manifold. The natural choice for such a form is the area form, $\omega = S \sin\theta d\phi \wedge d\theta$, 
where $(\phi,\theta)$ are the usual spherical coordinates. This allows us to introduce a Poisson bracket via the standard 
definition (turning $\mathbb{S}^2$ into a Poisson manifold),
\begin{align}\label{PsBcC}
\left\{f,g\right\} := (\omega^{-1})^{ij}(\partial_i f)(\partial_j g)\,.
\end{align}
Here, $f$ and $g$ are some smooth functions on phase space and $\omega^{-1}$ is inverse of the symplectic form $\omega$. 
Every Poisson bracket is a Lie bracket by definition. Calculating Eq. (\ref{PsBcC}) 
for components of the spin vector $\vec{S}$, we verify that they generate the $\mathfrak{su}(2)$ Lie algebra 
$\left\{S_i,S_j\right\} = \epsilon_{ijk}\!\:S^k$. Furthermore, integrating the symplectic form over the whole solid 
angle, we find that the area of phase space is
\begin{align}
\label{radius}
\int_{\!4\pi}\!\!\omega = 4\pi S < \infty\,,
\end{align}
which correctly has the dimension of the Planck constant. 
If we subsequently follow the topological quantization procedure, the finiteness of the phase 
space area leads to a specific, discrete spectrum of each of the three operators $\hat{S}_i$, as well as the 
spin square operator $\hat{\vec{S}}^2$. 

Let us now construct a continuous system of spins, whose phase space at each point of (physical) space is a two-sphere. 
Components of a continuous spin variable, $\vec{S}({\bf x}) = (S^x({\bf x}),S^y({\bf x}),S^z({\bf x}))$, are 
functions of a position vector ${\bf x}$ and satisfy the $\mathfrak{su}(2)$ algebra bracket in the distributional sense,
\begin{equation}\label{Salgebra}
\left\{S^i({\bf x}),S^j({\bf y})\right\} = \epsilon^{ijk}\!\:S_k({\bf x})\,\delta^{(3)}({\bf x - y})\,, 
\end{equation}
where $i,j,k \in \left\{x,y,z\right\}$ are the internal indices. Since the Poisson algebra, spanned by spin 
components, is three-dimensional, it is associated with one propagating degree of freedom and one Casimir invariant, 
$\vec{S}^2\! := \vec{S}\!\cdot\!\vec{S} \in {\rm const}$. To simplify the formalism, we may characterize the 
spin system by a parameter $S$ (spin magnitude), assumed to be a constant in space and defined as
\begin{equation}\label{norm}
S := ||\vec{S}|| = \sqrt{\vec{S} \cdot \vec{S}}\,.
\end{equation}
In the limit where $S$ is very large compared with all other scales, we will require that the 
ordinary scalar field theory is recovered [cf., (\ref{Hamiltonian1})]. Let us also emphasize that the spin 
magnitude is a scalar with respect to symmetry transformations [cf., the constraint \eqref{RvarphiRpiS} that we 
impose below] and that it appears in the map [Eqs. \eqref{Sx}--\eqref{Sz}] as a scaling factor. Thus, the value of $S$ 
for a given system is fixed up to a choice of units, i.e., its relation to other constants appearing in our further 
construction. This relation will be determined by the requirements that the Poisson algebra remains 
$\mathfrak{su}(2)$ and that the model of a spin system matches the Klein-Gordon field theory [see Eq.~(\ref{RvarphiRpiS}) 
and Eqs. \eqref{par1}--\eqref{par3}].

In principle, $S$ could become a dynamical quantity in a quite natural generalization of the 
discussed theory. For example, one might consider the quantization of $S^2$. The corresponding operator has a nontrivial 
spectrum, and its expectation value is determined by the characteristics of a quantum state, which is dynamical. 
However, in the present classical case, as noted above, $S$ is a Casimir of (the universal enveloping 
algebra of) the Poisson algebra and hence is fixed at all times by construction.

At the next step, we set the correspondence between the phase spaces of a spin system and a scalar field. We choose 
to do it by introducing the canonical parametrization of the sphere (at every point ${\bf x}$) in the following way:
\begin{align}
S^x &:= S \sqrt{1 - \frac{\pi^2_{\varphi}}{R^2_{\pi}}}\, \cos\left(\frac{\varphi}{R_{\varphi}}\right), \label{Sx} \\
S^y &:= S \sqrt{1 - \frac{\pi^2_{\varphi}}{R^2_{\pi}}}\, \sin\left(\frac{\varphi}{R_{\varphi}}\right), \label{Sy} \\
S^z &:= S \frac{\pi_{\varphi}}{R_{\pi}}\,, \label{Sz}  
\end{align}
where $\pi_{\varphi} \in [-R_{\pi},R_{\pi}]$ and $\varphi \in (-\pi R_{\varphi}, \pi R_{\varphi}]$. The latter 
coordinates are related to spherical ones $\theta,\phi$ via $\frac{\pi_{\varphi}}{R_{\pi}} = \cos\theta$, 
$\frac{\varphi}{R_{\varphi}} = \phi$ (up to trivial shifts of the ranges of angles). The fields $\varphi({\bf x})$ 
and $\pi_{\varphi}({\bf x})$ are then linearlike canonical variables [cf. \eqref{canonicalbracket}], in terms of 
which, we will be able to recover the ordinary scalar field, after choosing the appropriate Hamiltonian. 

Similar to spherical coordinates, $\varphi,\pi_\varphi$ obviously do not cover the whole sphere; i.e., $\varphi$ is 
ambiguously defined at the poles $\pi_\varphi = \pm R_{\pi}$. In order to extend the map [Eqs. \eqref{Sx}--\eqref{Sz}] 
everywhere, one has to make an analytic continuation. On the other hand, for the Hamiltonian [Eq. \eqref{Hamiltonian2}] 
that we will consider in this paper, the only phase space trajectories that approach the poles are the ones going 
along two pieces of the meridian $\varphi = \pm \sqrt{\frac{S}{m}} \frac{\pi}{2}$, where Eq. (\ref{Hamiltonian2}) 
vanishes (see Fig.~\ref{psts}). Thus, these two trajectories can be consistently joined into a single one. 
The Hamiltonian Eq. (\ref{Hamiltonian2}) is globally defined on $\mathbb{S}^2$ in terms of coordinates $S^x,S^y,S^z$ 
as Eq. (\ref{Hamiltonian0}), and the latter actually vanishes at the poles as well. Furthermore, the restricted range of 
applicability of $\varphi,\pi_\varphi$ does not pose a problem from the perspective of the considered spin-field 
correspondence, which is valid only in the regime of small field values -- or equivalently large spin magnitude -- far away from the 
poles (see the next section; in particular, Figs.~\ref{Sphere} and \ref{psts}).

The constants $R_{\varphi}$ and $R_{\pi}$ were introduced due to dimensional reasons. They play roles of parameters, 
which control the accuracy in recovering the standard form of the scalar field's Hamiltonian in the large $S$ limit. 
Moreover, as mentioned earlier, we expect only one independent Casimir invariant because the rank of the 
$\mathfrak{su}(2)$ algebra is two, and its dimension is three. Therefore, performing a canonical rescaling, one can 
eliminate $R_{\pi}$ and $R_{\varphi}$ in favor of $S$. It is worth noting that when we consider the general-relativistic 
perspective, the weights of the scalar $\varphi$ and the scalar density $\pi_{\varphi}$ will become relevant. Consequently, 
the parameters $R_{\pi}$ and $R_{\varphi}$ would also become a scalar and a scalar density, respectively.

In order to determine how the constants $R_{\varphi}$ and $R_{\pi}$ depend on $S$, we use the canonical bracket 
in the field formulation. The Poisson bracket is defined as usual,
\begin{align}\label{PoissonField}
\{A,B\}_{\varphi,\pi} := \int\! d^3z
\left(\frac{\delta A}{\delta \varphi({\bf z})} \frac{\delta B}{\delta\pi({\bf z})}
- \frac{\delta A}{\delta\pi({\bf z})} \frac{\delta B}{\delta\varphi({\bf z})}\right),
\end{align}
where $A := A[\varphi({\bf x}),\pi_{\varphi}({\bf x})]$ and $B := B[\varphi({\bf y}),\pi_{\varphi}({\bf y})]$ are arbitrary 
functionals. Computing the following relation, $\left\{S^z,S^x\right\}_{\varphi,\pi} = \frac{S}{R_{\varphi} R_{\pi}} S^y$, it is 
easy to see that to maintain consistency with the $\mathfrak{su}(2)$ algebra, the parameters have to be related via a simple equation
\begin{equation}\label{RvarphiRpiS}
R_{\varphi} R_{\pi} = \sqrt{q}\,S\,.
\end{equation}
This relation provides a constraint between the parameters $R_{\varphi}$, $R_{\pi}$ and the Casimir 
invariant $S$. In Eq. \eqref{par2}, we will choose a specific solution to this constraint such that our model matches the 
Klein-Gordon Hamiltonian in the low-energy limit. Since $R_{\varphi}$, $R_{\pi}/\sqrt{q}$ and $S$ have trivial 
Poisson brackets in this system, Eq. (\ref{RvarphiRpiS}) does not generate a flow that could leave the constraint surface.

Notice that we introduced above the quantity $q$, denoting the determinant of the spatial metric tensor $q_{ab}$ (which encodes 
invariance under spatial diffeomorphisms in the field formulation of the model) on the Cauchy hypersurface --- from the 
general-relativistic perspective, this would balance both sides of the identity. It would also modify the Levi-Civita symbol in Eq.
\eqref{Salgebra}, which would no longer be a tensor, but a totally antisymmetric tensor codensity (of weight $-1$),
\begin{align}\label{Levi-Civita}
\epsilon_{abc}:=\frac{1}{\sqrt{q}}\tilde{\epsilon}_{abc}\,.
\end{align}
Here, $\tilde{\epsilon}_{abc}$ is a tensor, while the indices $a,b,c,...$, which replaced $i,j,k,...$, emphasize that the 
considered space may be curved\footnote{The interesting property of the spin's system is the locality of the orientation 
of the spin vector's components, indicating directions in $\mathbb{R}^2$ with respect to the same point. In the single 
point model, associated with a rotationally invariant reference frame, the flat and curved coordinates are indistinguishable. 
Therefore, we do not need to modify the labeling of the internal coordinates in Eq. \eqref{Salgebra}. The contravariant or 
covariant position of the indices, however, is relevant --- it helps to control the proper weights of different objects. 
In consequence, this appears to specify the isotropic contribution to the minimal coupling between the gravitational 
field and the spin.}. The former set of indices labels the internal coordinates (in the spin formulation) curved accordingly 
to the spatial coordinates (in the field formulation) due to the dependence on the same metric, $q_{ab}$.

Provided the identity in Eq. \eqref{RvarphiRpiS} is satisfied, the algebra in Eq. (\ref{Salgebra}) is consistent with the assumption 
that the pair $\varphi$ and $\pi_{\varphi}$ satisfy the standard canonical bracket,
\begin{equation}\label{canonicalbracket}
\left\{\varphi({\bf x}),\pi_{\varphi}({\bf y})\right\}_{\varphi,\pi} = \delta^{(3)}({\bf x} - {\bf y})\,. 
\end{equation}
Let us also mention that we are interested in the cosmological application of the discussed spin-field correspondence. 
The standard form of the Poisson bracket will simplify the description of the cosmological perturbations, allowing one to 
perform a decomposition of the phase space into homogeneous and inhomogeneous parts in the standard way \cite{Mukhanov:1990me}.

It is necessary to stress that the parametrization [Eqs. \eqref{Sx}--\eqref{Sz}] has been chosen so that the $\varphi$ and 
$\pi_{\varphi}$ fields satisfy the canonical bracket  Eq. (\ref{canonicalbracket}). This is different from the spherical parametrization 
considered in Refs. \cite{Mielczarek:2016xql,Bilski:2017gic,Mielczarek:2017zoq}, which led to a modified form of the canonical 
relation between the field variables $\varphi$ and $\pi_{\varphi}$.

The canonical parametrization of $\mathfrak{su}(2)$ [Eqs. \eqref{Sx}--\eqref{Sz}] is technically advantageous due to the harmonic 
behavior of the variables in the vicinity of the classical minimum and the canonical relationship between the scalar field $\varphi$ 
and its momentum. On the other hand, its form might appear contrived. In order to motivate the form of this parametrization 
we note that it is the semiclassical limit of the well-known Holstein-Primakoff transformation \cite{Holstein:1940zp,HolsteinPrimakoff}. 
This transformation expresses $\mathfrak{su}(2)$ generators in terms of creation and annihilation operators, furnishing the crucial 
canonical structure. While it is not difficult to demonstrate the correspondence between the Holstein-Primakoff transformation and 
our canonical parametrization, the details of this semiclassical limit are too long to be included here and, therefore, were moved to 
Appendix~\ref{A}.

\section{Spin-field correspondence} \label{sec:III}
\begin{figure}[ht!]
\centering
\includegraphics[width=6cm,angle=0]{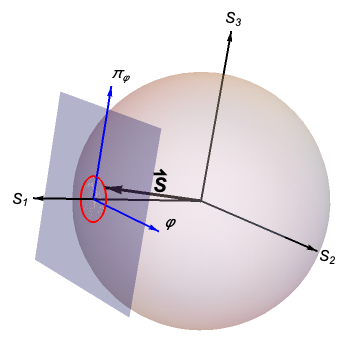}
\caption{Illustration of the precession of a spin vector $\vec{S}$ around the ground state 
$\vec{S} = (S,0,0)$, which corresponds to $(\varphi,\pi_\varphi) = (0,0)$. For the precession angle 
tending to zero, the dynamics is approximately described by harmonic oscillations at the $(\varphi,\pi_\varphi)$ 
phase space. In general, our parametrization maps a hemisphere into a nonsquare rectangle (cf., Fig.~\ref{psts}).}
\label{Sphere}
\end{figure}

\noindent
Let us consider the following Hamiltonian 
\begin{equation}\label{Hamiltonian0}
H := -\gamma \int d^3x S^x\,,
\end{equation}
where $\gamma$ is a certain constant. In condensed matter physics, it would be 
interpreted as the interaction term of a continuous distribution of internal magnetic moments (spins) with 
an external homogeneous magnetic field oriented along the $x$ axis. In the absence of other 
interactions, a spin at each point would precess around the ground state, $\vec{S} = (S,0,0)$.
Meanwhile, the same system, described in terms of the $\varphi$ and $\pi_\varphi$ variables, would 
correspond to a set of oscillators, which become harmonic when the precession angle tends to zero, 
which is pictorially represented in Fig.~\ref{Sphere}.

To make the spin-field correspondence evident, let us substitute the expression for $S^x$ given by (\ref{Sx}) into the Hamiltonian in Eq.
(\ref{Hamiltonian0}), finding,
\begin{align}
H &= -\gamma S \int d^3x \sqrt{1 - \frac{\pi^2_{\varphi}}{R^2_{\pi}}}\, \cos \left(\frac{\varphi}{R_{\varphi}}\right) \nonumber  \\ 
&= \int d^3x \left[-\gamma S + \left( \frac{\gamma S}{R^2_{\pi}} \right) \frac{\pi^2_{\varphi}}{2} + 
\left( \frac{\gamma S}{R^2_{\varphi}} \right) \frac{\varphi^2}{2}\right] \nonumber \\
&+ \mathcal{O}(\pi^4_{\varphi},\pi^2_{\varphi}\varphi^2,\varphi^4)\,. 
\label{Hamiltonian1}
\end{align}
Up to the constant factor $-\gamma S$, the leading order terms describe the
free homogeneous scalar field. From the mechanical perspective, the field is simply a continuous distribution of harmonic oscillators. 
Furthermore, the form of the Hamiltonian in Eq. (\ref{Hamiltonian0}) has been chosen in the way supporting the correspondence between
its ground state $\vec{S} = (S,0,0)$ and the field configuration $(\varphi, \pi_\varphi) = (0,0)$.  

The standard Klein-Gordon form of the Hamiltonian requires the following setting of parameters, $\frac{q \gamma S}{R^2_{\pi}} = 1$ 
and $\frac{\gamma S}{R^2_{\varphi}} = m^2$, where $m$ is the self-interaction constant -- the mass of $\varphi$, while $q$ balances 
the weights on both sides of the former identity\footnote{Notice that in Eq. \eqref{Hamiltonian1}, we did not consider the generally relativistic 
formulation of the field theory. Later, however, we are going to construct an extension to this formulation; therefore, any map linking the 
spin and field formulation, has to already include a proper scaling with respect to the metric tensor $q_{ab}$. Luckily, all the equations 
contributing to the map, which we discuss, being the Holstein-Primakoff transformation with a particular parametrization, involve only 
scalars or scalar densities and, being one of these two objects, fixed vector coordinates. Therefore, the proper scaling is provided only 
by appropriate powers of the determinant of $q_{ab}$.}. These two relations combined with the one in Eq. (\ref{RvarphiRpiS}) are equivalent 
to the set of three equations:
\begin{align}
\gamma &= m\,, \label{par1}\\
R_{\varphi} &= \sqrt{\frac{S}{m}}\,, \label{par2}\\ 
R_{\pi} &= \sqrt{q m S}\,. \label{par3}
\end{align}
The specification of parameters is in general not unique. This one, however, entails the homogeneous Klein-Gordon formulation of the spin 
wave representation related to the Holstein-Primakoff transformation of the Hamiltonian in Eq. \eqref{Hamiltonian0}, which describes interactions 
in a continuous distribution of spins in the presence of an external homogeneous magnetic field oriented along the $x$ axis. Let us emphasize 
that once the transformation in Eqs. \eqref{Sx}--\eqref{Sz}, associated with the parametrization \eqref{par1}--\eqref{par3}, is assumed, it fixes the 
meaning of the bosonic field's representation of the spin system. Consequently, it would also specify the quantum spin waves' kinematics and 
dynamics, whose solutions on the non-degenerate eigenstates of observables correspond to the excitations of a magnon quasiparticle. 
This allows one to study bosonic representations of spin systems, which are usually simpler generalizable and coupleable with other fields. 
Results of any analogous modification in the field formulation of a model are traceable then in the spin formulation.

The parametrization \eqref{par1}--\eqref{par3} applied to expression Eq. \eqref{Hamiltonian1} leads to the following form of the Hamiltonian:
\begin{align}
H &= -m S \int\! d^3x\ \sqrt{1 - \frac{\pi^2_{\varphi}}{S m}}\, \cos\left(\varphi \sqrt{\frac{m}{S}}\right) \label{Hamiltonian2} \nonumber  \\
&= -m S \int\! d^3x + \int\! d^3x \left(\frac{\pi^2_{\varphi}}{2} + m^2 \frac{\varphi^2}{2}\right) \nonumber \\
&+ \mathcal{O}(\pi^4_{\varphi},\pi^2_{\varphi}\varphi^2,\varphi^4)\,,
\end{align}
where we imposed the flat space constraint, $q_{ab}=\delta_{ab}$, simplifying the determinant of the metric tensor to $q=1$.
Notice that the first term in the second line diverges with the spacetime volume. This term, however, does not contribute to classical 
dynamics and, for convenience, can be regulated by performing an infinite subtraction (setting the vacuum energy to zero).

\begin{figure}[h]
\centering
\includegraphics[width=0.4\textwidth]{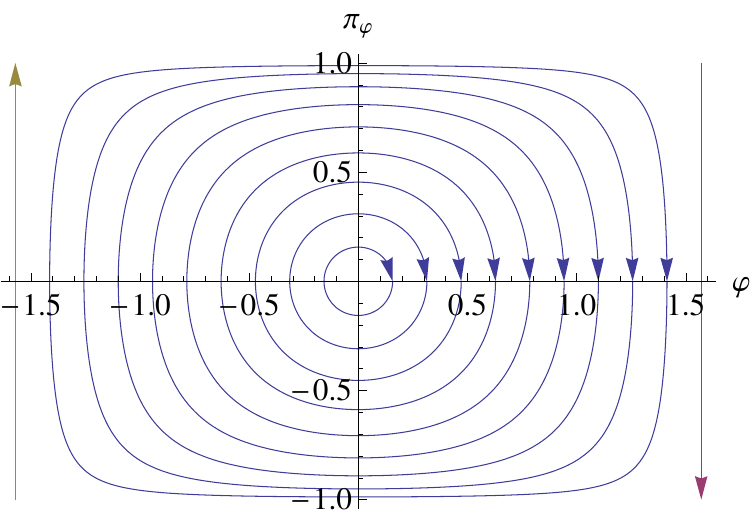}
\caption{A sample of phase space trajectories Eq. (\ref{hamsos}) on the hemisphere 
with $\varphi \in \left[-\frac{\pi}{2} \sqrt{\frac{S}{m}},\frac{\pi}{2} \sqrt{\frac{S}{m}}\right]$, 
for $S = 1$, $m = 1$; the arrows denote the direction of time.}
\label{psts}
\end{figure}

The evolution equations are calculated from the Hamiltonian Eq. (\ref{Hamiltonian2}) in the usual way, 
via $\dot\pi_\varphi = \{\varphi,H\}_{\varphi,\pi}$ and $\dot\varphi = \{\pi_\varphi,H\}_{\varphi,\pi}$,
\begin{align}\label{hameqs}
\dot\pi_\varphi &= -m \sqrt{m S - \pi_\varphi^2}\, \sin\left(\varphi \sqrt{\frac{m}{S}}\right)\,, \nonumber\\ 
\dot\varphi &= \frac{\sqrt{m S}\, \pi_\varphi}{\sqrt{m S - \pi_\varphi^2}} \cos\left(\varphi \sqrt{\frac{m}{S}}\right)\,.
\end{align}
Clearly, in the limit $S \rightarrow \infty$, the standard relations $\dot{\varphi} = \pi_{\varphi}$ and $\dot\pi_\varphi = -m^2 \varphi$ 
are correctly recovered. The solutions of the above equations for $\varphi \in \left(-\frac{\pi}{2} \sqrt{\frac{S}{m}},\frac{\pi}{2} \sqrt{\frac{S}{m}}\right)$ 
can be written as (cf., \cite{Trzesniewski:2017lpb}),
\begin{align}\label{hamsos}
\pi_\varphi(t) &= -C \sqrt{m S}\, \sin(m (t - t_0))\,, \nonumber\\ 
\varphi(t) &= \sqrt{\frac{S}{m}}\, \arcsin\left(\frac{C \cos(m (t - t_0))}{\sqrt{1 - C^2 \sin^2(m (t - t_0))}}\right)\,.
\end{align}
If we assume the initial condition for $\pi_\varphi$, $\pi_\varphi(t_0) = 0$, a constant $C = \sin(\sqrt{\frac{m}{S}}\, \varphi_0)$ 
encodes the other initial condition $\varphi(t_0) = \varphi_0 \in \left[0,\frac{\pi}{2} \sqrt{\frac{S}{m}}\right)$, equivalent to 
$\varphi(t_0 + \frac{\pi}{m}) = -\varphi_0 \in \left(-\frac{\pi}{2} \sqrt{\frac{S}{m}},0\right]$. The reason that we can restrict to 
setting an initial condition for $\varphi$ in the range $\left[0,\frac{\pi}{2} \sqrt{\frac{S}{m}}\right)$ is that all phase space trajectories 
$(\pi_\varphi,\varphi)$ are closed curves, as depicted in Fig.~\ref{psts}. Moreover, in the limit $\varphi_0 \to \frac{\pi}{2} \sqrt{\frac{S}{m}}$, 
we have $\varphi(t) = \pm\varphi_0|_{-\sqrt{m S} < \pi_\varphi < \sqrt{m S}}$. These are the only trajectories that 
reach the ill-defined boundaries $\pi_\varphi = \pm\sqrt{m S}$ (corresponding to poles of the sphere), and hence, their 
ending points can be pairwise identified in the unambiguous way. The formulae in Eq. (\ref{hamsos}) can be also analytically 
continued to the other hemisphere, leading to the mirror image. 

The first equation in Eq. (\ref{hameqs}) can be rewritten in the form:
\begin{equation}\label{phisol}
\frac{\sqrt{1 - \frac{\pi^2_{\varphi}}{m S}}}{\sqrt{m S} \cos\left(\varphi \sqrt{\frac{m}{S}}\right)} = 
\pm\frac{1}{\sqrt{\dot{\varphi}^2 + m S \cos^2\left(\varphi \sqrt{\frac{m}{S}}\right)}}\,.
\end{equation}
The $\pm$ sign corresponds to $\varphi \in {\cal F}_+$ or $\varphi \in {\cal F}_-$, respectively, where 
these two sectors of possible values of $\varphi$ (associated with two hemispheres) are
\begin{align}\label{sects}
&{\cal F}_+ := \sqrt{\frac{S}{m}}\, \left(-\frac{\pi}{2},\frac{\pi}{2}\right)\,, \\
&{\cal F}_- := \sqrt{\frac{S}{m}}\, \left( \left(-\pi,-\frac{\pi}{2}\right) \cup \left(\frac{\pi}{2},\pi\right] \right)\,,
\end{align}
while for $\varphi = \pm \sqrt{\frac{S}{m}} \frac{\pi}{2}$, we have $\dot\varphi = 0$. 
The sign ambiguity in Eq. (\ref{phisol}) has consequences when one performs the inverse Legendre 
transform to recover the Lagrangian. Namely, we have to define two variants of the Lagrangian,
\begin{align}\label{Lagrangian0}
L^\pm &:= \int\! d^3x\ \dot{\varphi} \pi_{\varphi} - H\ \vert_{\varphi \in {\cal F}_\pm} \nonumber \\
&= \pm\sqrt{m S} \int\! d^3x\ \sqrt{\dot{\varphi}^2 + m S \cos^2\left(\varphi \sqrt{\frac{m}{S}}\right)}
\end{align}
and it vanishes at the meridian $\varphi = \pm\sqrt{\frac{S}{m}} \frac{\pi}{2}$.

The two possible choices for the Lagrangian define dynamics related to two hemispheres of the spherical phase space. 
However, only $L^+$ allows us to recover the standard scalar field theory, corresponding to the large spin limit. This argument 
specifies the range of the possible values that the considered scalar field can take, $\varphi \sqrt{\frac{m}{S}} \in (-\pi/2,\pi/2)$, and it makes the $\varphi \rightarrow 0$ limit achievable. Consequently, in the expansion of $L^+$ around the small field's 
values and the related velocities,
\begin{align}
L^+ &= m S \int\! d^3x + \int\! d^3x\ \left(\frac{1}{2} \dot{\varphi}^2 - \frac{1}{2} m^2 \varphi^2 \right) \nonumber \\
&+ \mathcal{O}(\dot{\varphi}^4,\dot{\varphi}^2\varphi^2,\varphi^4)\,,
\label{Lagrangian1}
\end{align}
the standard form of the homogeneous scalar field's Lagrangian is recovered.
 
\section{Special relativistic extension} \label{sec:IV}
\noindent
In the previous section the spin-field correspondence was introduced, where the field side of this relation was constructed 
in the Hamiltonian formalism. The spin phase space was parametrized in the way leading to the spatially homogeneous 
Klein-Gordon-like form of the Hamiltonian in the large spin limit. It is, however, much easier to construct an invariant theory 
at the level of an action -- one just has to make sure that the action is a scalar. Difficulties appear when the Poisson bracket 
associated with a continuous system of spins is non canonical, and therefore derivation of the Lagrangian is neither straightforward, 
nor trivial. Another argument to focus on the field formalism is the well-known model of the Lorentz-invariant Klein-Gordon 
field, which is the standard formulation of this theory.

Before we begin generalizing the field side of our framework, let us mention how one could proceed with the spin-side. 
The standard (special) relativistic generalization of spin is introduced as follows (see, e.g., \cite{Misner:1973gn}). A given
system has the angular momentum tensor $J^{\mu\nu} = L^{\mu\nu} + S^{\mu\nu}$, where $L^{\mu\nu} = X^\mu P^\nu - X^\nu P^\mu$ 
is the orbital angular momentum, while the intrinsic part (i.e., spin) $S^{\mu\nu}$ can be expressed as 
$S^{\mu\nu} = \varepsilon^{\mu\nu\alpha\beta} u_\alpha S_\beta$. Consequently, the spin four-vector (known as Pauli-Luba\'{n}ski vector) 
is orthogonal to the four-velocity, $u_\alpha S^\alpha = 0$, and in the rest frame, it is simply $(S^\mu) = (0,\vec S)$. $M^{\mu\nu}$ 
and $S^{\mu\nu}$ become mixed under the action of Lorentz transformations unless $M^{\mu\nu} = 0$, as in our case. However, 
what we do in the current paper is quite different. We do not consider the relativistic generalization of a single spin but of a spin 
field, which is first represented by a scalar field and only then generalized.

Looking for the Lorentz-invariant analog of the correspondence defined by the transformation and parametrization in Eqs. \eqref{Sx}--\eqref{Sz} 
and Eqs. \eqref{par1}--\eqref{par3}, respectively, is the primary problem of this section. We are going to achieve our goal by modifying the Lagrangian 
obtained in Eq. (\ref{Lagrangian0}). This object is a functional of Lorentz scalars, and in the large spin limit, it takes the form of the homogeneous 
scalar field Lagrangian. Its generalization is then straightforward,
\begin{equation}
\dot{\varphi}^2\rightarrow -\eta^{\mu\nu} \partial_\mu\varphi \partial_\nu\varphi\,,
\end{equation}
where the Minkowski metric, $\eta^{\mu\nu} = \text{diag}(-1,1,1,1)$, was introduced. In consequence, Eq.~\eqref{Lagrangian0} 
leads to
\begin{align}
L^\pm &= \pm\sqrt{m S} \int\! d^3x\ \sqrt{-\eta^{\mu \nu} \partial_\mu\varphi \partial_\nu\varphi + m S \cos^2\left(\varphi \sqrt{\frac{m}{S}}\right)} \nonumber \\
&= \pm m S \int\! d^3x \pm \int\! d^3x\ \left( -\frac{1}{2} \eta^{\mu\nu} \partial_\mu\varphi \partial_\nu\varphi - \frac{1}{2} m^2 \varphi^2 \right) \nonumber \\
&+ \text{h.o.t.}
\label{Lagrangian2}
\end{align}
(from now on, the abbreviation h.o.t.\! denotes higher order terms). For the $+$ case, the leading order of the Lagrangian 
describes the standard Klein-Gordon scalar field.  The momentum canonically conjugate to $\varphi$ is given by the formula,
\begin{equation}\label{momentum_SR}
\pi_{\varphi} := \frac{\delta L^\pm}{\delta \dot{\varphi}} = \pm \frac{\sqrt{m S}\, \dot{\varphi}}{\sqrt{-\eta^{\mu\nu} \partial_\mu\varphi \partial_\nu\varphi + m S \cos^2\left(\varphi \sqrt{\frac{m}{S}}\right)}}\,.
\end{equation}
Performing the inverse Legendre transformation, we obtain the Hamiltonian,
\begin{widetext}
\begin{align}
H^\pm &= \mp\sqrt{m S} \int\! d^3x\ \sqrt{1 - \frac{\pi^2_{\varphi}}{m S}}\, \sqrt{-(\nabla \varphi)^2 
+ m S \cos^2\left(\varphi  \sqrt{\frac{m}{S}}\right)} \nonumber \\
&= \mp m S \int\! d^3x \pm \int\! d^3x\ \left( \frac{\pi^2_{\varphi}}{2} 
+ \frac{1}{2}(\nabla \varphi)^2 + m^2 \frac{\varphi^2}{2} \right)
+ \text{h.o.t.}\,, \label{Hamiltonian3}
\end{align}
\end{widetext}
which naturally vanishes at the meridian $\varphi \sqrt{\frac{m}{S}} = \pm \frac{\pi}{2}$.

The interesting consequence of the form of this Hamiltonian is that gradient of the field $\varphi$ is bounded, 
$|\nabla \varphi| \leq \sqrt{mS}$. Thinking of the gradient as the momentum, this relation allows us to 
put an upper bound on the energy carried by the scalar field waves in a finite volume. 

To get rid of the $\mp$ sign factor, one may move $\cos\left(\varphi \sqrt{\frac{m}{S}}\right)$ out of 
the square root, using the fact that different signs correspond to two hemispheres ${\cal F}_\pm$ 
and, respectively, positive or negative values of the $\cos\left(\varphi \sqrt{\frac{m}{S}}\right)$ function. 
As the result, the Hamiltonian becomes
\begin{widetext}
\begin{align}\label{Hamiltonian4}
H = -m S \int\! d^3x\ \sqrt{1 - \frac{\pi^2_{\varphi}}{m S}}\, \sqrt{1 - \frac{(\nabla \varphi)^2}{m S 
\cos^2\left(\varphi \sqrt{\frac{m}{S}}\right)}}\, \cos\left(\varphi \sqrt{\frac{m}{S}}\right).
\end{align}
\end{widetext}

Using the relations,
\begin{equation}
\cos\left(\varphi \sqrt{\frac{m}{S}}\right) = \frac{S^x}{\sqrt{(S^x)^2 + (S^y)^2}}
\end{equation} 
and
\begin{equation}
\varphi = \sqrt{\frac{S}{m}}\, \arctan\left(\frac{S^y}{S^x}\right)\,,
\end{equation}
we can then express the relativistic Hamiltonian in Eq. (\ref{Hamiltonian4}) in terms of the spin variables,
\begin{equation}\label{HI}
H_{\vec S} = -m \int\! d^3x\ S^x \sqrt{1 - \frac{1}{m^2} \frac{\left(S^y \nabla S^x - S^x \nabla S^y\right)^2}{(S^x)^2 [(S^x)^2 + (S^y)^2]}}\,.
\end{equation}

In the large mass $m$ limit and in a vicinity of the ground state (for which $S^x \approx S$ and $S^y \approx 0$), 
the Hamiltonian can be approximated by the following expression
\begin{equation}\label{smallgrad}
H_{\vec S} \approx -\frac{1}{m S} \int\! d^3x\ \left[-\frac{1}{2}(\nabla S^y)^2 + m^2 S S^x\right].
\end{equation}

In the context of condensed matter physics, this would be interpreted as a continuous Ising model 
coupled to a constant, transversal external magnetic field \cite{Pfeuty}. Interestingly, the model 
is not only of theoretical interest and finds an important application in adiabatic quantum computing 
\cite{Kadowaki}.  Therefore, quantized Hamiltonian Eq. (\ref{HI}) may allow one to study special relativistic 
generalisations of the quantum annealing process. Furthermore, in the following section, we will
extend the Hamiltonian Eq. (\ref{HI}) to the general relativistic case.  

\section{General relativistic extension} \label{sec:V}

\noindent Before we proceed, let us note that there are different views on what is the 
correct way to include the intrinsic angular momentum (i.e., spin) in gravitation. Namely, it may be the 
standard Einsteinian general relativity \cite{Bailey:1975ly} or the metric-affine gauge theory 
\cite{Blagojevic:2012gn}, the simplest example of which is Einstein-Cartan(-Sciama-Kibble) theory 
\cite{Trautman:2006ey}. In the latter case, the intrinsic angular momentum turns out to be a source of 
the nonvanishing torsion. 
We choose to consider in this paper the standard (torsion-less) general relativity.

Having established the canonical formulation of the bosonic wave representation of the spin system, characterized 
by the Lorentz symmetry, the most natural way to proceed is to look for an extension of the symmetry to the full generally 
relativistic case. This construction can be done in much the same way as in the special relativistic case; hence, except 
instead of using the Minkowski metric, as in Eq. \eqref{Lagrangian2}, we are going to use the general one, $g_{\mu\nu}$. 
Consequently, the general relativistic extension of Eq. \eqref{Lagrangian0} involves the following replacing,
\begin{equation}
\dot{\varphi}^2 \rightarrow -g^{\mu\nu} \partial_\mu\varphi \partial_\nu\varphi\,,
\end{equation}
Our goal is to formulate this scalar field's system in the canonical setting, and therefore, it is most convenient to decompose 
the metric tensor into ADM variables,
\begin{equation}\label{gmunu}
g_{\mu\nu} = \left(\begin{array}{cc} -N^2 + q_{ab} N^a N^b & N_a \\ N_b & q_{ab} \end{array}\right)\,.
\end{equation}
Here, $N$ denotes the lapse function, $N^a$ the shift vector, and $q_{ab} = g_{ab}$ the spatial metric, where $a,b,... = 1,2,3$ 
label spatial coordinates. Consequently, the volume element takes the form $\sqrt{-g} = N \sqrt{q}$, where $g=\det(g_{\mu\nu})$ 
and $q=\det(q_{ab})$. The inverse of Eq. (\ref{gmunu}) is given by the matrix
\begin{equation}\label{gmunuinv}
g^{\mu\nu} = \left(\begin{array}{cc} -\frac{1}{N^2} & \frac{N^b}{N^2} \\ \frac{N^a}{N^2} & q^{ab} - \frac{N^a N^b}{N^2} \end{array}\right),
\end{equation}
Using this object, we can express the kinetic term as,
\begin{align}
-g^{\mu\nu} \partial_\mu\varphi \partial_\nu\varphi &= \frac{1}{N^2} \dot{\varphi}^2 - \frac{2}{N^2} \dot{\varphi} N^a \partial_a\varphi \nonumber\\
&- \left(q^{ab} - \frac{N^a N^b}{N^2}\right) \partial_a\varphi \partial_b\varphi \nonumber\\
&= \frac{1}{N^2} (\dot{\varphi} - N^a \partial_a\varphi)^2 - q^{ab} \partial_a\varphi \partial_b\varphi\,.
\end{align}
This form of the kinetic term in the field formulation of the action allows one to begin the canonical analysis and, in particular, to derive the momentum of the field $\varphi$.

As a next step, we postulate the fully relativistic generalization of the action based on the Lagrangian in Eq. (\ref{Lagrangian2}),
\begin{widetext}
\begin{align}
S^\pm_\varphi &=\pm \sqrt{m S} \int\! d^4x\ \sqrt{q} N \sqrt{-g^{\mu \nu} \partial_\mu\varphi \partial_\nu\varphi + m S \cos^2\left(\varphi \sqrt{\frac{m}{S}}\right)} \\
&= \pm m S \int d^4x\sqrt{q}N \pm \int\! d^4x\ \sqrt{q} N \left(-\frac{1}{2} g^{\mu\nu} \partial_\mu\varphi \partial_\nu\varphi - \frac{1}{2} m^2 \varphi^2 \right) + \text{h.o.t.}.
\label{ActionCov}
\end{align} 
\end{widetext}
The result, as expected, took an explicitly invariant form. The transformation of the model back to the spin-formulation can 
be done analogously to the procedure in the previous section.

Dynamical analysis of the general-relativistic extension of a field theory has to involve gravity. This is needed because 
the extension is realized via the minimal coupling of the field with the metric tensor. In order to make the spacetime geometry 
dynamical, we define the minimal coupling of the theories in the standard way -- at the level of their actions, constructing the following one:
\begin{equation}\label{SFA}
S := S_{\rm G} + S_\varphi^\pm = \int\! dt \left(L_{\rm G} + L_\varphi^\pm\right)\,.
\end{equation} 
The formulation of our model allows us also to easily add other elements to the theory such as a cosmological constant or other types of bosonic fields. 

To complete the construction of the generally relativistic spin-field correspondence theory, we now need to perform the Legendre transformation of Eq. \eqref{SFA}. The direct calculation gives the momentum canonically conjugate to $\varphi$,
\begin{align}\label{momentum_GR}
\pi_{\varphi} :=&\;\frac{\delta L^\pm_\varphi}{\delta \dot{\varphi}} \nonumber \\
=&\;\pm \frac{ \sqrt{m S}\, \frac{\sqrt{q}}{N}(\dot{\varphi} - N^a \partial_a\varphi)}{\sqrt{-g^{\mu\nu} \partial_\mu\varphi \partial_\nu\varphi + m S \cos^2\left(\varphi \sqrt{\frac{m}{S}}\right)}}.
\end{align}
In the large spin limit ($S \rightarrow \infty$) of the $+$ case, we correctly recover $\pi_{\varphi} = \frac{\sqrt{q}}{N} (\dot{\varphi} - N^a \partial_a\varphi)$. The matter Hamiltonian can now be written as
\begin{align}\label{matHam}
H_\varphi^\pm[N,N^a] = \int\! d^3x\ \dot{\varphi} \pi_{\varphi} - L_\varphi^\pm = H_\varphi^\pm[N] + D_\varphi[N^a]\,, 
\end{align}
where we introduced the field's diffeomorphism constraint,
\begin{equation}\label{Dphi}
D_\varphi[N^a] = \int\! d^3x\ N^a \pi_{\varphi} \partial_a\varphi\,.
\end{equation}
It took the same form as in the case of the ordinary Klein-Gordon field. Finally, the matter Hamiltonian constraint reads,
\begin{widetext}
\begin{equation}\label{HamCov}
H_\varphi^\pm[N] = \mp\sqrt{m S} \int\! d^3x\ N \sqrt{q}\, \sqrt{1 - \frac{\pi^2_{\varphi}}{q m S}}\, \sqrt{-q^{ab} \partial_a\varphi \partial_b\varphi 
+ m S \cos^2\left(\varphi \sqrt{\frac{m}{S}}\right)}\,.
\end{equation}
\end{widetext}
Taking the limit where the field excitations are small compared with the scale set by $S$, we correctly recover the Hamiltonian 
constraint of the self-interacting scalar field minimally coupled to Einsteinian gravity.

\subsection{Hypersurface deformation algebra}
\noindent
Even though we constructed the Hamiltonian in Eq. (\ref{matHam}) using the general relativistic approach, it is useful to 
perform an independent check that this Hamiltonian is indeed covariant. This can be done
verifying that our spin-field contributions lead to the unmodified constraint algebra. 
In other words, our constraints should satisfy the hypersurface deformation algebra. Taking a different 
point of view, we can think of Eqs. \eqref{HamCov} and \eqref{Dphi} as an ansatz for the contributions 
from the matter component to the gravitational constraint and then verify that they close the algebra.

The total constraints are sums of the gravity and matter contributions,
\begin{align}
D[N^a] &= D_{\text{G}}[N^a] + D_{\varphi}[N^a]\,, \\
H[N] &= H_{\text{G}}[N] + H_{\varphi}[N]\,.
\end{align}
We will first check whether the following identity holds: $\left\{H[N],H[M]\right\} = D[q^{ab} \left( N \partial_b M - M \partial_bN \right)]$.
To this end, we simplify the problem, collecting together the Hamiltonian contributions from different fields,
\begin{equation}
\label{algebra1}
\begin{split}
\left\{H_{g}[N] + H_{\varphi}[N],H_{g}[M] + H_{\varphi}[M]\right\} &= \\
\left\{H_{g}[N],H_{g}[M]\right\} + \left\{H_{\varphi}[N],H_{\varphi}[M]\right\} &.
\end{split}
\end{equation}
The cross terms canceled out because the minimal coupling of the matter field to gravity
in the ADM formalism is given by coupling only with the spatial 
metric (and not with the gravitational momenta). Consequently, no integration by parts has to 
be performed when computing the cross terms and their sum is proportional to $N M - M N = 0$.

The first term in the second line of Eq. \eqref{algebra1} is standard because we did not 
consider any modification of the gravitational constraint. Therefore, we only 
need to compute the second term, which by the direct calculation is found to be,
\begin{equation}
\left\{H_{\varphi}[N],H_{\varphi}[M]\right\} = D_{\varphi}[q^{ab} \left( N\partial_b M - M \partial_bN \right)]\,.
\end{equation}
This is the result that we expected -- the contribution to the diffeomorphism 
constraint from the matter field takes the standard form for the Klein-Gordon field, despite our modifications. 

The calculation of the bracket between the diffeomorphism constraint and the 
Hamiltonian constraint is a little more subtle. Let the object
$\mathcal{H} = \mathcal{H}[q_{ab},\pi_{ab},\varphi,\pi_{\varphi}]$ denote the Hamiltonian 
densit; then the following relation holds:
\begin{widetext}
\begin{equation}
\begin{split}
\left\{\mathcal{H},D_{\text{G}} + D_{\varphi}\right\} &= \int\! d^3x\, \Big[\frac{\delta \mathcal{H}}{\delta \varphi}\frac{\delta \left(D_{\varphi} + D_{\text{G}}\right)}{\delta \pi_{\varphi}}
- \frac{\delta (D_{\varphi} + D_{\text{G}})}{\delta \varphi}\frac{\delta \mathcal{H}}{\delta \pi_{\varphi}} + \frac{\delta \mathcal{H}}{\delta q_{a b}} \frac{\delta \left(D_{\varphi} + D_{\text{G}}\right)}{\delta \pi^{ab}} - \frac{\delta (D_{\varphi} + D_{\text{G}})}{\delta q^{ab}}\frac{\delta \mathcal{H}}{\delta \pi^{ab}}\Big]\\
&= \int\! d^3x\, \Big[\frac{\delta \mathcal{H}}{\delta \varphi}\left\{\varphi,D_{\varphi}\right\} + \left\{\pi_{\varphi},D_{\varphi}\right\} \frac{\delta \mathcal{H}}{\delta \pi_{\varphi}}
+\frac{\delta \mathcal{H}}{\delta q_{a b}}\left\{q_{ab},D_{\text{G}}\right\} + \left\{\pi^{ab},D_{\text{G}}\right\}\frac{\delta \mathcal{H}}{\delta \pi^{ab}}\Big]\\
&= \int\! d^3x\, \Big[\frac{\delta \mathcal{H}}{\delta \varphi}\mathcal{L}_{\vec{N}}\varphi + \frac{\delta \mathcal{H}}{\delta \pi_{\varphi}}\mathcal{L}_{\vec{N}}\pi_{\varphi}
+ \frac{\delta \mathcal{H}}{\delta q_{a b}}\mathcal{L}_{\vec{N}}q_{ab} + \frac{\delta \mathcal{H}}{\delta \pi^{ab}}\mathcal{L}_{\vec{N}}\pi^{ab}\Big]\\
&= \mathcal{L}_{\vec{N}}\mathcal{H}\,.
\end{split}
\end{equation}
\end{widetext}
We can use this intermediate result to compute the bracket between the Hamiltonian 
constraint and the diffeomorphism one, obtaining
\begin{widetext}
\begin{equation}
\begin{split}
\left\{H[N],D_{G}[N^i] + D_{\varphi}[N^i]\right\} &=
\int\! d^3x\, \left(\mathcal{H}\left\{N,D[N^i]\right\} + N\left\{\mathcal{H},D[N^i]\right\}\right) \\
&= \int\! d^3x N\, \mathcal{L}_{\vec{N}}\mathcal{H}
= -\int\! d^3x\, \mathcal{H} \mathcal{L}_{\vec{N}}N 
= -H[\mathcal{L}_{\vec{N}}N].
\end{split}
\end{equation}
\end{widetext}
It is worth mentioning that we derived this result almost without any effort due to the unmodified matter contribution to 
the diffeomorphism constraint.

Collecting all the resulting Poisson 
brackets together, we obtain the following list:
\begin{align}
\left\{D[N^a],D[M^a]\right\} &= D[\mathcal{L}_{\vec{N}}M^a]\,, \\ 
\left\{H[N],D[N^a]\right\} &= -H[\mathcal{L}_{\vec{N}}N]\,, \\
\left\{H[N],H[M]\right\} &= D[q^{ab}( N\partial_b M - M \partial_bN)]\,.  
\end{align}
We can then conclude this section with a remark that our spin-field matter contribution
indeed leads to a generally relativistic invariant theory when coupled to a dynamical,
possibly curved background.
 
\subsection{General relativistic spin-field correspondence}
\noindent
Similar to what we did in the special-relativistic case, we begin with absorbing the sign factor
by moving the function $\cos^2\left(\varphi \sqrt{\frac{m}{S}}\right)$ in (\ref{HamCov}) out of 
the square root, which leads to the following Hamiltonian:
\begin{widetext}
\begin{equation}\label{HamCov2}
H_\varphi[N] = -m S \int d^3 x N \sqrt{q} \sqrt{1 - \frac{\pi^2_{\varphi}}{q m S}} 
\sqrt{1 - \frac{q^{ab} \partial_a\varphi \partial_b\varphi}{m S \cos^2\left(\varphi \sqrt{\frac{m}{S}}\right)}} \cos\left(\varphi \sqrt{\frac{m}{S}}\right)\,.
\end{equation}
\end{widetext}

As a next step, we reexpress this Hamiltonian in terms of spin variables, obtaining
\begin{widetext}
\begin{equation}\label{HamCovNonPert}
H_{\vec{S}}[N] = -m \int\! d^3x \sqrt{q}\, N\ S^x \sqrt{1-\frac{1}{m^2}q^{ab}\partial_{a}\left(\text{arcsinh}{\left(\frac{S^y}{S^x}\right)}\right)
\partial_b\left(\text{arcsinh}{\left(\frac{S^y}{S^x}\right)}\right)}\,.
\end{equation}
\end{widetext}
Notice that this result puts an upper bound on the magnitude of the gradient of $\text{arcsinh}\left(\frac{S^y}{S^x}\right)$. 
It is also worth mentioning that deriving the expression above, we took the advantage of the correct construction of the 
map in [cf. Eqs. \eqref{Sx}--\eqref{Sz}], in which the weights of $\pi_{\varphi}$ are balanced with the ones of $R_{\varphi}$ in the 
parametrization [Eqs. \eqref{par1}--\eqref{par3}]. To emphasize the importance of the correctness of the transformation's construction, 
let us recall the $\mathfrak{su}(2)$ algebra from Eq. \eqref{Salgebra}, expressing it in the explicitly covariant notation,
\begin{equation}\label{algebra}
\left\{S_a(x),S_b(y)\right\} = \epsilon_{abc}\!\:S^c(x)\,\delta^{(3)}(x - y)\,.
\end{equation}
Let us also recall the implicit coupling to the metric tensor of two objects in the formula above, $S_a=q_{ab}S^b$ and 
$\epsilon_{abc}=\tilde{\epsilon}_{abc}/\sqrt{q}$, where $S^b$ is the spin vector in the general coordinates, while 
$\tilde{\epsilon}_{abc}$ is the Levi-Civita tensor.

Finally, the spin formulation of the diffeomorphism constraint in the generally relativistic framework [given in Eq. \eqref{Dphi}] reads,
\begin{widetext}
\begin{equation}\label{Dphi_GR_S}
D_{\vec{S}}[N^a] = \frac{1}{S} \int\! d^3x\ \sqrt{q}\, N^a \frac{S^z}{(S^x)^2 + (S^y)^2}
\left(S^x \partial_a S^y - S^y \partial_a S^x \right)\,.
\end{equation}
\end{widetext}

We reexpressed the matter contributions to the constraints in terms of 
spin variables but, although we have changed coordinates, the calculations done in the previous 
section with the hypersurface deformation algebra still hold. This is due to the independence
of the overall Poisson structure of the choice of coordinates.
 
\section{Dirac-Born-Infeld Theory perspective} \label{sec:VI}
\noindent
We will now demonstrate that our model can be mapped to a Dirac-Born-Infeld (DBI) \cite{Leigh:1989jq,Silverstein:2003hf,Alishahiha:2004eh} 
model via the appropriate field redefinition. Let us begin moving the cosine square term out of the square root the action in Eq. (\ref{ActionCov}) 
so that the sign factor is absorbed -- analogously as in Sec.~\ref{sec:V}; this time, however, at the level of the action,
\begin{widetext}
\begin{align}
S_\varphi^\pm &= \pm\sqrt{m S} \int\! d^4x \sqrt{q}\, N\ \sqrt{-g^{\mu\nu} \partial_\mu\varphi \partial_\nu\varphi
+ m S \cos^2{\left(\varphi \sqrt{\frac{m}{S}}\right)}} \nonumber \\
&= \int\! d^4x \sqrt{q}\, N\ m S \cos\left(\varphi \sqrt{\frac{m}{S}}\right)
\sqrt{1 - \frac{g^{\mu\nu} \partial_\mu\varphi \partial_\nu\varphi}{m S \cos^2\left(\varphi \sqrt{\frac{m}{S}}\right)}}\,.
\label{Spmphi}
\end{align}
\end{widetext}
The resulting expression has similar form to the free Dirac-Born-Infeld (DBI) action for a scalar field $\xi$,
\begin{equation}\label{DBIaction}
S_{\rm DBI} := \int\! d^4x \sqrt{q}\, N\ \frac{1}{f(\xi)} \sqrt{1 - f(\xi)\, g^{\mu\nu} \partial_\mu\xi \partial_\nu\xi}\,.
\end{equation}
Here, $f(\xi)$ is a functional, which ,in the case of the D3-brane inspired origin of the DBI action, is a warp 
factor of the AdS-like throat \cite{Alishahiha:2004eh}.

To find the relation between Eqs. \eqref{Spmphi} and (\ref{DBIaction}), let us redefine the field variable in the former 
expression, setting $\varphi = G(\xi)$. The form of the functional $G(\xi)$ depends on whether the sector 
$\varphi \in {\cal F}_+$ or $\varphi \in {\cal F}_-$ [see Eq. (\ref{sects})] is considered. The first sector corresponds 
to the non-negative sign of $\cos{\left(\varphi \sqrt{\frac{m}{S}}\right)}$, while in the second one, to the negative 
sign. To distinguish these situations, we introduce the additional labeling of $G(\xi)$; i.e., the functional 
related to the first sector is going to be denoted by $G_+(\xi)$, while to the second one, $G_-(\xi)$.

Applying the change of variables, $\varphi = G(\xi)$, to the action in Eq. (\ref{Spmphi}), we obtain,
\begin{widetext}
\begin{equation}\label{Splus2}
S_\varphi = \int\! d^4x \sqrt{q}\, N\ m S \cos{\left(G(\xi) \sqrt{\frac{m}{S}}\right)}
\sqrt{1 - \frac{(G'(\xi))^2 g^{\mu\nu} \partial_\mu\xi \partial_\nu\xi}{m S \cos^2\left(G(\xi) \sqrt{\frac{m}{S}}\right)}}\,.
\end{equation}
\end{widetext}
This suggests to impose the following constraints on $G_\pm(\xi)$,
\begin{equation}\label{Gcon}
\left(\frac{dG_\pm(\xi)}{d\xi}\right)^2 = \pm\cos\left(G_\pm(\xi) \sqrt{\frac{m}{S}}\right)\,.
\end{equation}
This leads the DBI-like form of the action,
\begin{equation} \label{DBIlikeAction}
S_{\xi}= \int\! d^4x \sqrt{q}\, N\ \frac{1}{f(\xi)} \sqrt{1 - |f(\xi)|\, g^{\mu\nu} \partial_\mu\xi \partial_\nu\xi}\,,
\end{equation}
where the functional $f(\xi)$ is given by the expression,
\begin{equation}\label{fxifunction}
f(\xi) = \frac{1}{m S \cos{\left(G(\xi)\sqrt{\frac{m}{S}}\right)}}\,.
\end{equation}
The latter is positive in the range $\varphi \in {\cal F}_+$, negative for 
$\varphi \in {\cal F}_-$, and diverges when $\varphi \sqrt{\frac{m}{S}} = \pm\frac{\pi}{2}$ 
(the action vanishes in the last case, since $f(\xi)$ also appears in the denominator). 
The action [Eq. \eqref{DBIlikeAction}] matches with the DBI action [Eq. \eqref{DBIaction}]
only in the ${\cal F}_+$ sector. The absolute value $|f|$ ensures that the argument 
$|f(\xi)|\, g^{\mu\nu} \partial_\mu\xi \partial_\nu\xi$ is always non-negative and the square 
root is real valued. This observation will have essential consequences when applying
this model to cosmology, in particular, during the inflation stage, which will be
discussed later, in Sec.~\ref{sec:VII}.

\begin{figure}[h!]
\includegraphics[scale=0.35]{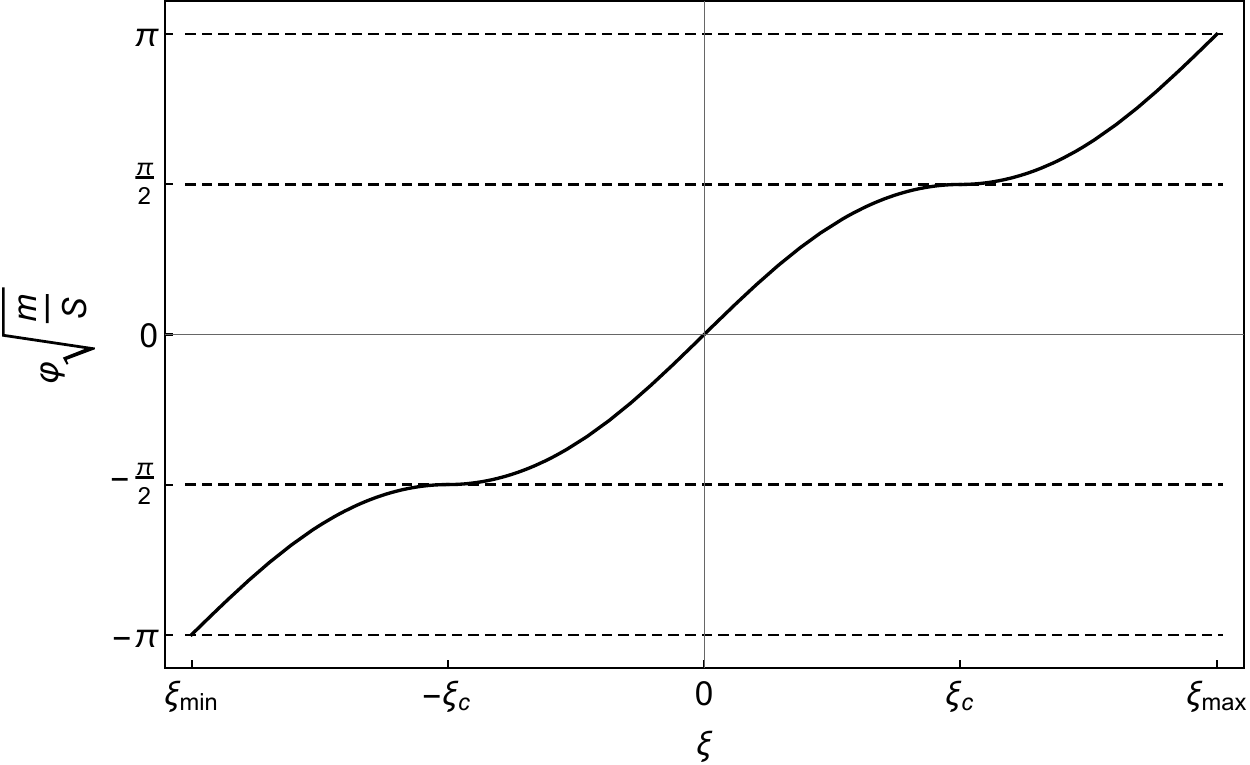}
\caption{Plot of the $\varphi = G(\xi)$ function composed of two branches: 
$G_+(\xi)$ for $\xi \in [-\xi_c,\xi_c ]$ and $G_-(\xi)$ for $\xi \in (\xi_{\text{min}},-\xi_c) \cup (\xi_c,\xi_{\text{max}}]$. 
We see that this function is increasing monotonically and covers the whole range of variability of $\varphi$.}
\label{GfunctionPlot}
\end{figure}

Let us first consider the positive branch. Choosing $G_+(0) = 0$ and $G'_+(0) = 1$ so that in a vicinity of $\xi = 0$ we have $\varphi \approx \xi$, we find that the solution 
to the Eq. (\ref{Gcon}) takes the form,
\begin{align}\label{Gplus}
&G_+(\xi) = 2 \sqrt{\frac{S}{m}}\, {\rm am}\left(\sqrt{\frac{m}{S}}\, \frac{\xi}{2} \Big\vert 2\right), \nonumber  \\
&|\xi| \leq 2 \sqrt{\frac{S}{m}}\, F\left(\frac{\pi}{4} \Big\vert 2\right) := \xi_c\,,
\end{align}
where ${\rm am}(u | n)$ is the Jacobi amplitude function, and 
$F(x | n) := \int_0^x \frac{d\theta}{\sqrt{1 - n \sin^2\theta}}$ is the 
elliptic integral of the first kind. They are known to satisfy the relation if $x = {\rm am}(u | n)$, then 
$u = F(x | n)$. Moreover, the critical value of $|\xi| =: \xi_c$ corresponds to   
the limiting values of $|\varphi| = \frac{\pi}{2}\sqrt{\frac{S}{m}}$.

In the negative branch, assuming the boundary 
conditions $G_-(\pm \xi_c) = \pm\frac{\pi}{2} \sqrt{\frac{S}{m}}$, we find the following solution to Eq. (\ref{Gcon}): 
\begin{widetext}
\begin{equation}\label{Gminus}
G_-(\xi) = \left\{ \begin{array}{ccc} 2 \sqrt{\frac{S}{m}}\, {\rm am}\left(\frac{i}{2} 
\left(\xi - \left(1 + i\right) \xi_c \right) \sqrt{\frac{m}{S}} \Big\vert
2\right) & \text{for} & \xi \in (\xi_c,\xi_{\text{max}}] \\ 
2 \sqrt{\frac{S}{m}}\, {\rm am}\left(\frac{i}{2} \left(\xi + \left(1 + i\right) \xi_c \right) \sqrt{\frac{m}{S}} \Big\vert
2\right) & \text{for} & \xi \in (\xi_{\text{min}},-\xi_c)\end{array} \right. .
\end{equation}
\end{widetext}
Despite the presence of imaginary factors $i$, the above functions are real valued. The minimal 
and maximal values of $\xi$ are
\begin{align}
\xi_{\text{min}} &= -\xi_c + 2i \left( F\left(\frac{\pi}{4} \Big\vert  2\right) - F\left(\frac{\pi}{2} \Big\vert 2\right) \right) \sqrt{\frac{S}{m}}\,, \\
\xi_{\text{max}} &= \xi_c - 2i \left( F\left(\frac{\pi}{4} \Big\vert  2\right) - F\left(\frac{\pi}{2} \Big\vert 2\right) \right) \sqrt{\frac{S}{m}}\,. 
\end{align}

The $\varphi = G(\xi)$ function in the full range of variability of its argument (covering 
the $G_+(\xi)$ and $G_-(\xi)$ branches) is plotted in Fig.~\ref{GfunctionPlot}.

In the low-energy limit, in which gradients are small, the DBI-like action [Eq. (\ref{DBIlikeAction})] takes 
the approximate form 
\begin{equation}
S_{\xi} = \int\! d^4x \sqrt{q}\, N \left(\mp \frac{1}{2} g^{\mu\nu} \partial_\mu\xi \partial_\nu\xi + \frac{1}{f(\xi)}\right) + \text{h.o.t.}\,,
\end{equation}
where the $\mp$ sign refers to the ${\cal F}_+$ or ${\cal F}_-$ sector, respectively.

In such a case, the functional $f(\xi)$ plays the role of the effective potential,
\begin{equation}
U(\xi) := -\frac{1}{f(\xi)} \approx -m S + \frac{1}{2} m^2 \xi^2 - \frac{m^3}{8 S} \xi^4 + \mathcal{O}(\xi^6)\,.
\end{equation}
Then the mass of the field $\xi$ is indeed represented by the scalar $m$. The quartic self interaction term is of the order 
$\mathcal{O}(1/S)$ as the neglected higher order kinetic term.

Finally, performing the large $S$ expansion of the action in Eq. (\ref{DBIaction}) with the functional $f(\xi)$ given by the 
expression in Eq. (\ref{fxifunction}), we obtain,
\begin{widetext}
\begin{align}
S_\xi &= m S \int\! d^4x \sqrt{q} N
+ \int\! d^4x \sqrt{q}\, N \left(\mp\frac{1}{2} g^{\mu\nu} \partial_\mu\xi \partial_\nu\xi - \frac{1}{2} m^2 \xi^2\right) \nonumber\\
&+ \frac{1}{2m S} \int\! d^4x \sqrt{q}\, N \left(-\frac{1}{2} g^{\mu\nu} \partial_\mu\xi \partial_\nu\xi - \frac{1}{2} m^2\right) 
\left(+\frac{1}{2} g^{\mu\nu} \partial_\mu\xi \partial_\nu\xi - \frac{1}{2} m^2\right) + \mathcal{O}(1/S^2)\,.
\end{align}
\end{widetext}

This action gives the following equations of motion for $\xi$,
\begin{widetext}
\begin{equation}
\begin{split}
0&=\mp\frac{1}{\sqrt{-g}}\partial_{\nu}\left[\sqrt{-g} g^{\mu \nu}\partial_{\mu}\xi\right]+m^2 \xi\\
&+ \frac{1}{m S} \Bigg[ \frac{1}{\sqrt{-g}} \partial_{\nu}\left[\sqrt{-g} \left(\frac{1}{2} g^{\mu\nu} \partial_{\mu}\xi \partial_{\nu}\xi + \frac{1}{2}m^2 \xi^2\right)g^{\alpha\beta} \partial_{\alpha}\xi\right] - \left[\frac{1}{2} g^{\mu\nu} \partial_{\mu}\xi \partial_{\nu}\xi + \frac{1}{2} m^2 \xi^2\right] m^2 \xi \Bigg]
+ \mathcal{O}(S^{-2})\,,
\end{split}
\end{equation}
\end{widetext}
corresponding to the sectors, ${\cal F}_+$ and ${\cal F}_-$, respectively. To unify the notation, the identity, 
$N\sqrt{q}=\sqrt{-g}$, was used. The first two terms contribute to the standard wave equation for the 
Klein-Gordon on curved spacetime, while the terms of order $\mathcal{O}(S^{-1})$ specify the nonlinear corrections.

\section{Cosmological implications} \label{sec:VII}

\noindent
What would a continuous spin system embedded into spacetime imply on its dynamics? In the previous section, 
a parallel between the general relativistic spin action and the DBI-type action has been established, which permits one
to make some predictions toward this direction, being of particular relevance in the cosmological context.

This is because, since a few decades, works on string theory led to a renewed interest concerning the DBI action 
due to its link with the $D$-brane models \cite{Leigh:1989jq}. Imposing the DBI action [Eq. (\ref{DBIaction})] to be real 
requires the square root $\sqrt{1 - f(\xi)\, g^{\mu\nu} \partial_\mu\xi \partial_\nu\xi}$ to be real valued. Under this 
condition -- and considering negative $f$ function, an upper bound on the scalar field velocity $|\dot{\xi}|$ is predicted. 
This upper bound is responsible for the so-called $D$-cceleration mechanism and can be shown to introduce
naturally a slow-roll inflation \cite{Silverstein:2003hf}, with promising predictions on the non-Gausiannity of the power 
spectrum of the CMB \cite{Alishahiha:2004eh}. This sections aims to investigate briefly if such cosmological 
features are also implied by our model, while we keep a more detailed analysis for forthcoming publication.

\subsection{Stress-energy tensor}

\noindent
The stress-energy tensor for the scalar field $\xi$ with the action [Eq. (\ref{DBIlikeAction})] is
\begin{align}
T^{\mu\nu} &:= \frac{2}{\sqrt{-g}} \frac{\delta S_{\xi}}{\delta g_{\mu\nu}} \nonumber\\
&= \frac{g^{\mu\nu}(1 - |f|\, \partial_\delta\xi \partial^\delta\xi)
+ |f| \partial^\mu\xi \partial^\nu\xi}{f \sqrt{1 - |f|\, \partial_\delta\xi \partial^\delta\xi}}\,.
\label{StressEnergyTensorXi}
\end{align}
From here, energy density ($\rho$) and pressure ($p$) of the scalar field can be extracted 
using the standard expression, valid for the thermal equilibrium case,
\begin{equation}
T^{\mu\nu} = (\rho + p ) u^\mu u^\nu + p g^{\mu\nu},
\label{StressEnergyTensor}
\end{equation} 
where $u^\mu$ is a four-velocity, satisfying the normalization condition $g_{\mu\nu} u^\mu u^\nu = -1$.

Because we are interested in cosmological consequences, let us from now on specify the metric
to be that of isotropic, homogeneous, and flat Friedmann-Lema\^{i}tre-Robertson-Walker (FLRW) 
models with the time gauge fixed by choosing the lapse function to be $N=1$. In this case, 
we have  $g_{\mu\nu}=(-1,a^2 \delta_{ij})$, $g^{\mu\nu} = (-1, \delta^{ij}/a^2)$, where $a$ denotes 
the scale factor. For this metric, in the comoving reference frame, the well normalized four-velocity  
vector takes the form $u^\mu = (1,0,0,0)$. In consequence, only diagonal elements of the 
stress-energy tensor [Eq. (\ref{StressEnergyTensor})] survive, leading to the relations: $\rho = T^{00}$ and $p = \frac{1}{3} g_{ij} T^{ij}$.

In consequence, for the isotropic and homogeneous background geometry, the stress-energy tensor 
of the (test) field $\xi$, given by Eq.~(\ref{StressEnergyTensorXi}), leads to the following expressions 
for energy density and pressure: 
\begin{equation}
\begin{split}
\rho &= -\frac{\alpha}{f}\,, \\
p &= \frac{1}{3 a^2}\, {\rm sgn}(f) \alpha (\nabla\xi)^2 + \frac{1}{\alpha f}\,,
\end{split}
\label{EnergyPressure}
\end{equation}
and
\begin{equation}
\alpha(\xi, \dot{\xi}) := \frac{1}{\sqrt{1 - |f(\xi)|\, g^{\mu\nu} \partial_\mu\xi \partial_\nu\xi}}\,.
\end{equation}
We emphasize that, because of the sign difference, this function does not correspond to the 
$\gamma$ Lorentz-like factor for $D$-branes in general. 

In the large spin limit when $\xi \rightarrow 0$, we obtain, 
\begin{align}
\rho &= \frac{1}{2}\dot{\xi}^2 - \frac{1}{2a^2} (\nabla \xi)^2 - m S + \frac{1}{2}m^2 \xi^2 + \mathcal{O}(1/S)\,, \\
p &= \frac{1}{2}\dot{\xi}^2 - \frac{1}{6a^2}(\nabla \xi)^2 + m S - \frac{1}{2}m^2 \xi^2 + \mathcal{O}(1/S)\,.
\end{align} 

The leading order terms are in agreement with the standard expressions for the Klein-Gordon scalar 
field. However, because of the constant contribution $mS$, the large spin limit ($S\rightarrow \infty$)
leads to divergences. The role of this constant is not specified at this level. One possibility to avoid the unphysical
behavior is to subtract the $mS$ term from the action  [(\ref{DBIlikeAction})] such that the new 
renormalised version is
\begin{align}
\tilde{S}_{\xi} &= \int\! d^4x \sqrt{q}\, N\  \frac{1}{f(\xi)} \sqrt{1 - |f(\xi)|\, g^{\mu\nu} \partial_\mu\xi \partial_\nu\xi} \nonumber \\
&- m S \int\! d^4x \sqrt{q}\, N\,.
\end{align}

\subsection{Cosmological evolution}

\noindent
In the considered case of the FLRW cosmology, the dynamics of a universe is entirely described 
by the Friedmann equations,
\begin{equation}
\begin{split}
H^2 &:= \left( \frac{\dot{a}}{a}\right)^2 = \frac{8\pi G}{3}\rho, \\
\dot{H} + H^2 &:= \frac{\ddot{a}}{a} = -\frac{4\pi G}{3} \left(\rho + 3p \right),
\end{split}
\label{FriedmannEqns}
\end{equation}
where $H$ is the Hubble factor, and $\ddot{a}/a$ is quantifying acceleration rate of expansion.

Expressing the equation of state in the form
\begin{equation}
p = w \rho,
\end{equation}
where $w$ is a function being a constant for the special case of barotropic fluid, it is convenient 
to discriminate between different types of evolution of a  universe. Namely, it follows from the 
second Friedmann equation [Eq. (\ref{FriedmannEqns})] that for a positive energy density,
\begin{equation}
\begin{split}
w &> -1/3 \Longrightarrow \ddot{a} < 0\,, \\
w &< -1/3 \Longrightarrow \ddot{a} > 0\,, \\
w &= -1/3 \Longrightarrow \ddot{a} = 0\,,
\end{split}
\end{equation}
while a negative energy density would imply change of sign for the acceleration in 
the expressions above. However, the latter case is physical only if appropriate modifications to 
the first Friedmann equation [Eq. \eqref{FriedmannEqns}] are present, ensuring that $H^2 \geq 0$.
 
Employing Eq. \eqref{EnergyPressure}, the homogeneous contributions from the 
scalar field $\xi$ to energy density and pressure are
\begin{equation}
\begin{split}
\rho &= - \frac{\alpha}{f}\,, \\
p &= \frac{1}{f\alpha}\,.
\end{split}
\end{equation}
One can immediately notice that the two cases where the energy density 
is positive or negative correspond to the two hemispheres of the phase space where, 
respectively,  $\varphi \in {\cal F}_-$ or $\varphi \in {\cal F}_+$.

We remind that, in Sec.~\ref{sec:III}, the Lagrangian of our model was shown to match 
with that of a scalar field in the large spin limit for the second case $\varphi \in {\cal F}_+$. 
We will, however, consider both possibilities in the following analysis.

More broadly, on both hemispheres, the parameter $w$ takes the form,
\begin{equation}
w = -\frac{1}{\alpha^2} = -1 - |f(\xi)| \dot{\xi}^2 \leq -1\,,
\label{wParameter}
\end{equation}
where in the considered FLRW case, $\alpha$ reduces to
\begin{equation}
\alpha(\xi, \dot{\xi}) := \frac{1}{\sqrt{1 + |f(\xi)| \dot{\xi}^2}}\,.  
\end{equation}
It is interesting to notice that in the Euclidean case (i.e., after the Wick rotation), 
the $\alpha$ function transforms precisely to the $\gamma$ Lorentz-like factor in DBI theory.   

While Eq. \eqref{wParameter} has similar functional form as in the usual DBI cosmology 
\cite{Silverstein:2003hf}, here, we have $\alpha \leq 1$ always. This ensures that $w\leq - 1$, 
and, in consequence the expansion of a universe is continually accelerated if $\rho>0$.
This behavior is similar to the one know from the case of \emph{phantom cosmologies} 
\cite{Caldwell:1999ew}. Furthermore, since $w<-1/3$, the more a universe expands, 
the more $\rho$ increases. In consequence, the more $\rho$ increases, the faster a 
universe expands, thus engendering internal inflation as soon as $\dot{a}>0$. The $D$-cceleration 
leading to slow-roll inflation is, therefore, excluded by our model.

On the other hand, assuming now that $f$ is positive ($\varphi \in \mathcal{F}_+$) 
such that the energy density is negative definite, a similar reasoning leads to the conclusion that 
the expansion is decelerated, which is, by definition, incompatible with an inflationary scenario.

One could, however, argue that the flatness and horizon problems can still be solved 
without the need for an inflationary phase. The other common solution requires a recollapse 
phase and refers to an ekpyrotic or cyclic universe \cite{Lehners:2008vx}.

Alternatively, both problems are solved through the emission of tachyacoustic perturbations, 
which also happen to be scale invariant \cite{Bessada:2009ns}. One can indeed observe that 
the speed of adiabatic waves for our model is larger than the speed of light in vacuum 
\cite{Garriga:1999vw, Alishahiha:2004eh},
\begin{equation}
c_s^2= \frac{1}{\alpha^2} =1+ |f(\xi)| \dot{\xi}^2 \geq 1.
\end{equation}
It is important to notice that these tachyacoustic perturbations for $k$-essence models \cite{ArmendarizPicon:1999rj} 
do not violate causality \cite{Babichev:2007dw} (see \cite{Bessada:2009ns} for an application of the 
proof to the DBI case).

Closing the discussion on the inflationary phase, the obtained DBI-like action may imply broader 
phenomenology and be considered as a candidate for dark energy \cite{Martin:2008xw}. Furthermore, 
worth stressing is also the relation to the Chaplygin gas models and tachyon condensate.   
Another important property is that the DBI-type $k$-essence models, to which our model belongs, 
have a unique property from the viewpoint of the problem of introducing time in gravity, by virtue 
of the Brown-Kuchar mechanism \cite{Thiemann:2006up}.   

\section{Summary} \label{sec:VIII}

\noindent In this article, the general-relativistic model of a continuous spin system has been 
constructed. It has been derived as a generalization of a system of spins
(magnetic moments) precessing in a constant magnetic field. An essential step 
was the application of the semiclassical version of the Holstein-Primakoff transformation, 
which allowed us to relate phase space of a spin field with phase space of a 
scalar field, so that the Poisson bracket on the latter remains canonical. The transformation is an example of a general procedure introduced 
recently in the context of nonlinear field space theories (NFSTs).     

The construction discussed in this article is not unique, and the method can be used 
to obtain other theories of spin fields on curved spacetimes. However, the considered case 
is special because of a few reasons. First, the investigated model reduces to the 
relativistic massive scalar field theory (Klein-Gordon field) in the large spin limit. Second, 
in the large mass limit and in the vicinity of a ground state, the model reduces to the 
continuous Ising model coupled to an external constant transversal magnetic field. 
Third, the model is equivalent to a concrete realisation of the DBI-type $k$-essence 
scalar field theory, with the form of the $f(\xi)$ function predicted by the model.

The third point is especially interesting since it indicates that there is a certain 
relation between spin fields and the Dirac-Born-Infeld theory. The spin field is, in turn, 
an example of NFST with the compact phase space. Actually, one of the motivations
behind proposing the NFST program was to impose constraints on the field 
values in the spirit of the original Born-Infeld theory. However, in the case of NFST, 
this is done by introducing nonlinearity to the field phase space. The results of our 
investigations confirm that some compact phase space realisations of NFST may be  
equivalent to the DBI-type scalar field theories. This also shows a possible connection 
between compact phase spaces and string theory, in the context of which, the DBI 
models have been considered in the recent literature. Furthermore, reduction to 
the DBI-type $k$-essence opens a possibility of making phenomenological predictions, 
especially in cosmology \cite{Alishahiha:2004eh}, which has been preliminarily explored 
here.

The approach introduced in this paper provides a consistent method of coupling a spin 
field to gravity, which may be of relevance not only at the classical but also at the quantum level. 
While one possibility given by our framework is coupling a spin field to quantum gravity, 
not less interesting is the analysis of quantum spin systems on curved backgrounds. Therefore, 
the considered model and the whole framework may find application in the domain of 
quantum many-body systems on curved spacetimes and the relativistic quantum information 
theory \cite{RQI}. This may be relevant, e.g., in theoretical description of such astrophysical 
objects as neutron stars, quark stars, or white dwarfs, where both quantum and gravitational 
effects (but not quantum-gravitational) are relevant. For this purpose, quantization of the 
spin field considered in the paper has to be performed, which is a challenge that interested 
readers are encouraged to take. 

\vspace{-0.25cm}

\section*{Acknowledgements}

\noindent
We thank Martin Bojowald for his comments on this work.
The research has been supported by the Sonata Bis Grant No. DEC-2017/26/E/ST2/00763 
of the National Science Centre Poland.
The work was partially supported by the National Natural Science Foundation of China with Grants No. 11675145 and No. 11975203.

\appendix
\section{semiclassical Holstein-Primakoff transformation}\label{A}
\noindent 
The Holstein-Primakoff transformation \cite{Holstein:1940zp} expresses 
the spin operators [generating the $\mathfrak{su}(2)$ algebra] in terms of the bosonic creation and annihilation operators. Restricting our interest to 
specific irreducible representations amounts to a truncation of the infinite-dimensional Hilbert 
space to a finite one. The transformation has the form (in the Cartan-Weyl basis $\hat S^\pm := \hat S^x \pm i \hat S^y$, $\hat S^z$),
\begin{equation}\label{quantum_algebra}
\begin{split}
\hat{S}^z &= \hbar \left(s - \hat{a}^{\dagger} \hat{a}\right)\,, \\
\hat{S}^{+} &= \hbar \sqrt{2s}\, \sqrt{1 - \frac{\hat{a}^{\dagger} \hat{a}}{2s}}\, \hat{a}\,, \\
\hat{S}^{-} &= \hbar \sqrt{2s}\, \hat{a}^{\dagger} \sqrt{1 - \frac{\hat{a}^{\dagger} \hat{a}}{2s}}\,,
\end{split}
\end{equation}
where the ladder operators satisfy the bosonic commutator, $\left[\hat{a},\hat{a}^{\dagger}\right] = 1$. 
We compute the semiclassical limit  of this algebra in the canonical manner, according to \cite{BOJOWALD_2006}. 
The relevant quantities in this limit are the expectation values of the considered operators, $\langle \hat{S}^{(z,\pm)} \rangle$.

The expectation value of the commutator entails the form of the Poisson bracket,
\begin{equation}\label{PsBc}
\left\{\langle \hat{A}\rangle,\langle \hat{B}\rangle\right\} = \frac{1}{i \hbar} \left\langle\left[\hat{A},\hat{B}\right]\right\rangle,
\end{equation}
so that, in particular, $\left\{a,a^{\dagger}\right\} = -\frac{i}{\hbar}$, where $a \equiv \langle \hat{a}\rangle$ and $a^{\dagger} \equiv \langle \hat{a}^{\dagger}\rangle$. 
Defining $S := \hbar s$ and $N := \langle\hat{a}^{\dagger}\rangle \langle\hat{a}\rangle$, 
we obtain the semiclassical expressions for spin variables in terms of bosonic variables,
\begin{equation}
\begin{split}
S^z &= S - \hbar N + O(\hbar^2)\,, \\
S^{+} &= \sqrt{\hbar}\, a\sqrt{2S - \hbar N} + O(\hbar^2)\,, \\
S^{-} &= \sqrt{\hbar}\, a^{\dagger}\sqrt{2S - \hbar N} + O(\hbar^2)\,.
\end{split}
\end{equation}
The correction terms come from the quantum fluctuations and from ordering ambiguities, which can be neglected in the semiclassical regime. 
Generators of the $\mathfrak{su}(2)$ algebra in the standard basis (i.e., $S^x$, $S^y$, $S^z$) are then given by,
\begin{equation}\label{class_map}
\begin{split}
S^z &= S - \hbar N\,, \\
S^x &= \frac{\sqrt{\hbar}}{2} \left(a + a^{\dagger}\right) \sqrt{2S - \hbar N}\,, \\
S^y &= \frac{\sqrt{\hbar}}{2i} \left(a - a^{\dagger}\right) \sqrt{2S - \hbar N}\,.
\end{split}
\end{equation}
They satisfy the $\mathfrak{su}(2)$ Lie algebra bracket, given by the Poisson bracket [Eq. (\ref{PsBc})] we just defined. 

At the next step, we introduce an action variable, $J := \hbar N$, which is canonically conjugate to some angle variable, 
$\theta$. The canonical structure implies that both $a$ and $a^{\dagger}$ depend on $\theta$, and the form of this dependence can be determined 
by considering the following brackets,
\begin{equation}
\begin{split}
\left\{a,J\right\} &= \frac{\partial a}{\partial \theta} = -i a\,, \\
\left\{a^{\dagger},J\right\} &= \frac{\partial a^{\dagger}}{\partial \theta} = i a^{\dagger}\,.
\end{split}
\end{equation}
The solution of the above system is
\begin{equation}
\begin{split}
a &= \sqrt{\frac{J}{\hbar}}\, {\rm e}^{-i \theta}\,, \\
a^{\dagger} &= \sqrt{\frac{J}{\hbar}}\, {\rm e}^{i \theta}\,,
\end{split}
\end{equation}
where $\left\{\theta,J\right\} = 1$, and the constants of integration were fixed by requiring $a^{\dagger} a = N$. 
Expressing the formulae in Eq. \eqref{class_map} in terms of the canonical variables, we ultimately obtain
\begin{equation}
\begin{split}
S^z &= S - J\,, \\
S^x &= \cos{(\theta)} \sqrt{J (2S - J)}\,, \\
S^y &= -\sin{(\theta)} \sqrt{J (2S - J)}\,.
\end{split}
\end{equation}

Last, to recover the parametrization selected in this paper, we perform the subsequent linear canonical transformation
\begin{equation}
\begin{split}
\varphi &= -R_{\varphi}\theta\,, \\
\pi_{\varphi} &= R_{\pi} - \frac{J}{R_{\varphi}}\,.
\end{split}
\end{equation}
Recalling the relation [Eq. \eqref{RvarphiRpiS}] and either moving the tensor codensity encoded in $1/\sqrt{q}$ into the Levi-Civita 
symbol [compare with the expression \eqref{Levi-Civita}] or setting the flat space limit, restricting to $q = 1$, and simplifying the 
relation \eqref{RvarphiRpiS} to $R_{\varphi} R_{\pi} = S$, we obtain the parametrization postulated in Eqs. \eqref{Sx}--\eqref{Sz}. This confirms that our  parametrization is the semiclassical analog of the Holstein-Primakoff transformation.

\end{document}